\documentclass[12pt]{iopart}

\usepackage{graphicx}
\usepackage{amsfonts,amsgen,amsbsy,amssymb}
\usepackage{iopams}
 \usepackage{cite}
\usepackage{filecontents}
 \usepackage{longtable}
\usepackage{enumitem}
\usepackage{lscape}
\usepackage{caption}
\usepackage{tabularx}

\begin{document}

\title[Chemotaxis Guided Amoeboid Cell Swimming]{Mathematical Modeling of Chemotaxis Guided Amoeboid Cell Swimming}

\author{Qixuan Wang$^{1,2,*}$ and Hao Wu$^3$}

\address{$^1$ Department of Mathematics, University of California, Riverside, CA, USA}
\address{$^2$ Interdisciplinary Center for Quantitative Modeling in Biology, University of California, Riverside, CA, USA}
\address{$^3$ Department of Polymer Science and Engineering, University of Massachusetts, Amherst, MA, USA}
\address{$^*$  Author to whom any correspondence should be addressed.}

\ead{qixuanw@ucr.edu}
\vspace{10pt}
\begin{indented}
\item[]December 2020
\end{indented}

\begin{abstract} 
Cells and microorganisms adopt various strategies to migrate in response to different environmental stimuli. 
To date, many modeling research has focused on the crawling-based \textit{Dictyostelium discoideum} (Dd) cells migration induced by chemotaxis,
yet recent experimental results reveal that even without adhesion or contact to a substrate, Dd cells can still
swim to follow chemoattractant signals. In this paper, we develop a modeling framework to investigate
the chemotaxis induced amoeboid cell swimming dynamics. A minimal swimming system consists of one
deformable Dd amoeboid cell and a dilute suspension of bacteria, and the bacteria produce chemoattractant signals that
attract the Dd cell. We use the \textit{mathematical amoeba model} to generate Dd cell deformation and solve the
resulting low Reynolds number flows, and use a moving mesh based finite volume method to solve 
the reaction-diffusion-convection equation. Using the computational model, we show that chemotaxis guides a 
swimming Dd cell to follow and catch bacteria, while on the other hand, bacterial rheotaxis may help the bacteria
to escape from the predator Dd cell.
\end{abstract}

\vspace{2pc}

\section{Introduction}

Cell migration, an integrated molecular process involving biochemical cascades intercorrelated with external chemical and mechanical stimuli, continues throughout the life span of many organisms \cite{Delanoe2008Changes}. 
Different microorganisms adopt various propulsion mechanisms and directed locomotion strategies for searching for food / running from predators. For example, individual cells and micro-organisms
such as \textit{C. reinhardtii} and spermatozoa find food by a combination of taxis and kinesis using a flagellated or ciliated mode of swimming \cite{Sokolov2010Swimming,Riedel2005Self,Lauga2006Swimming,Rafai2010Effective}. Other cell migration processes use the highly motile amoeboid mode, whose underlying molecular mechanisms have been extensively studied \cite{Manahan2004Chemoattractant}. Amoeboid (e.g., \textit{Dictyostelium discoideum} or leukocytes) cell migration relies on the generation, protrusion, and sometimes even travel of either pseudopodia or blebs \cite{howe2013amoebae}. As for strategies for directed locomotion, many flagellated bacteria (e.g., \textit{E. coli}, \textit{S. marcescens}, and \textit{V. alginolyticus}) and even some eukaryotic organisms such as the green algae \textit{C. reinhardtii} adopt a run- and-tumble type of motility \cite{Bray2000Cell}. During the run stage, the bacterium performs a more or less linear motion, while during the tumble stage it performs a highly erratic motion that produces little translocation but reorients the cell, thereby generating a random direction for the next run \cite{Stewart1987molecular,Block1982Impulse,Berg1990bacterial}. On the other hand, amoeboid cells detect extracellular chemical and mechanical signal gradients via membrane receptors, and these trigger signal transduction cascades that produce intracellular signals. Small differences in the extracellular signal are amplified into large end-to-end intracellular differences that control the motile machinery of the cell and thereby determine cell polarization and sites of pseudopod or bleb generation \cite{Paluch2005Cortical,Tinevez2009Role,Kay2002Chemotaxis,Kessin2001Dictyostelium,Bonner2009Social,Devreotes1988Chemotaxis,Bonner2015Cellular}.

Due to their genetic, biochemical and cell-biological tractability, the social amoebae -- \textit{Dictyostelium discoideum} (Dd) have been a
microorganism of choice to study basic processes in morphogenesis, including cell-cell chemical signaling, signal transduction, and cell
motility \cite{Manahan2004Chemoattractant,Meinhardt1999Orientation,Neilson2011Modeling,Neilson2011Chemotaxis,Hecht2011Activated}. 
Crawling-based chemotaxis-driven Dd migration, at both individual and collective levels, has been well
studied via both models and experiments
\cite{Tang1995Excitation,Dallon1997Discrete,Palsson2000Model,Dallon2004How,Khamviwath2013Continuum,Cheng2016Model,Bretschneider2016Progress,Dallon1998Continuum,Wilhelm2007Magnetic,Janmey2007Cell,Riviere2007Signaling,Decave2003Shear,Holmes2012Comparison}. 
To date, amoeboid cell migration and taxis are generally studied as the cells crawl on various solid substrates, relying on pseudopods attaching to the substrate. Recently, it was discovered that Dd cells can occasionally detach from the substrate and stay completely free in suspension for a few minutes before they slowly sink; during the free suspension stage, cells continue to form pseudopods that convert to rear-ward moving bumps, thereby propelling the cell through the surrounding fluid in a totally adhesion-free fashion \cite{van2011amoeboid}. Also, a mutant of Dd, sadA, which attaches poorly to a substrate, appears nevertheless to migrate normally and does so with an enhanced speed \cite{barry2010dictyostelium}. In the experiments, cells actively swam to a point source of cAMP, compared to no directed motion when the cAMP source is absent \cite{barry2010dictyostelium}.
Furthermore, a similar adhesion-independent swimming model involving large-scale shape deformation of the cell body may be adopted by other cells, in particular, traditionally well known crawling cells: for example, human neutrophils swim to a chemoattractant fMLP (formyl-methionylleucyl-phenylalanine) source at a speed similar to that of cells migrating on a glass coverslip under similar conditions \cite{barry2010dictyostelium}. Most recently and equally striking, cytokine can induce Drosophila fat body cells to actively swim to wounds in an adhesion- independent motility mode associated with actomyosin-driven, peristaltic cell shape deformations that initiate from the cortex of the cell center and extend to the rear of the cell, propelling them forward. These waves occur constantly within fat body cells in unwounded pupae and become highly directed with respect to a wound. Once at the wound, fat body cells start to form lamellipodia that extend around the wound margin, assist hemocytes to clear the wound of cell debris, tightly seal the epithelial wound gap, and release antimicrobial peptides to fight wound infection \cite{Franz2018Fat}.

Inspired by these recent experimental discoveries of amoeboid mode of swimming – in the strict sense of adhesion-independent cell-fluid interaction that involves large-scale of cell shape deformations, it is timely to conduct a modeling study on chemotaxis driven Dd swimming that allows the coupling of signaling dynamics and biohydrodynamics. In recent years, several models for single cell amoeboid swimming have emerged,
many focus on exploring the fluid-structure interaction in the system and how the amoeboid style of shape deformations lead to swimming in various viscous fluid environments
\cite{wang2016computational,Wu2015Amoeboid,Wu2016Amoebid,Aoun2020Amoeboid,Dalal2020Amoeboid,Farutin2013Amoeboid,Bouffanais2010hydrodynamics},
some also consider the underlying membrane protein kinetics that regulate the excitable dynamics of the the cell membrane deformations
\cite{Campbell2017Computational,Campbell2020Computational}. In this paper, we develop a 
model that includes a deforming Dd amoeboid cell and a group of bacteria, where the amoeboid cell swims following chemoattractant 
signal produced by the bacteria. The model is a minimal one that couples the chemotaxis dynamics and the hydrodynamic effects.
The paper is structured as follows. In section \ref{Sec.Model} we introduce the model setup, where the active bacterium motions are
modeled by a random walk model (section \ref{Sec.BacMot}), the chemotaxis signaling dynamics is numerically solved using
the finite volume method in a moving mesh (section \ref{Sec.SignDyn}), the Dd amoeboid cell shape deformations and the resulting 
fluid dynamics are modeled and solved using an established complex analysis technique (section \ref{Sec.CellDef-Swim}) 
\cite{shapere1987self,shapere1989geometry,avron2004optimal,Bouffanais2010hydrodynamics}.
Numerical results are presented and discussed in section \ref{Sec.Results}, where we first discussed the chemotaxis
guided amoeboid swimming with one bacterium (section \ref{Sec.ChemoDd}), then how bacterial rheotaxis could help
the bacteria escape from the predator Dd cell (section \ref{Sec.RheoBac}), finally chemotaxis guided amoeboid swimming with
a dilute suspension of bacteria (section \ref{Sec.MultiBac}).

 
\section{Modeling framework}\label{Sec.Model}

In this section we will discuss the development of the model, including the bacterial motions
(section~\ref{Sec.BacMot}), the chemotaxis dynamics (section~\ref{Sec.SignDyn}), the deforming 
Dd cell and the resulting fluid dynamics (section~\ref{Sec.CellDef-Swim}).
We consider a system consisting of a Dd amoeboid cell and bacteria in low Reynolds number incompressible Newtonian fluid.
For simplicity of modeling and computation, we will consider a 2D system. Many of the chemotaxis induced Dd cell swimming experiments
are performed in containers sufficiently large so as to avoid influences from the walls \cite{barry2010dictyostelium,van2011amoeboid}.
In this paper, we also consider a ``large tank" modeling system, where the fluid mechanics resulted mainly from the swimming deformable Dd cells
is obtained using the \textit{mathematical amoeba model} \cite{shapere1987self,shapere1989geometry,avron2004optimal,Bouffanais2010hydrodynamics,wang2016computational} 
approach, which provides the solution in a 2D infinite fluid domain (section~\ref{Sec.CellDef-Swim}); on the other
hand, the signaling dynamics is modeled using a moving-mesh based reaction-diffusion-convection (RDC) PDE model (section~\ref{Sec.SignDyn}), 
where we assume a finite but sufficiently large computation domain for the RDC system. Such a coupled modeling framework
allows us to efficiently study the dynamics of the system with relatively low computational costs, assuming the swimming Dd and bacteria 
are all away from the computational boundary of the RDC submodel.


\subsection{Bacterium motions.}\label{Sec.BacMot}

We consider \textit{Escherichia coli} (\textit{E. coli}) as a representative bacteria model. \textit{E. coli}'s typical movement strategy 
is well known as run-and-tumble: an \textit{E. coli} can either rotate its flagella counterclockwise resulting in a directed 
straight ``run", or rebundle its flagella by rotating them clockwise resulting in a ``tumble" which reorients the cell without significant change of location
\cite{Stewart1987molecular,Block1982Impulse}.
The \textit{E. coli} constantly switch between the run and tumble modes. The mean run interval is reported to be about 1 sec
in the absence of chemotaxis, and the mean tumble interval is about 0.1 sec, and both are distributed exponentially 
\cite{Berg1990bacterial}. In our model system, we will first consider one amoeboid Dd cell with
one \textit{E. coli}, due to the small size of a \textit{E. coli} (length $\sim 2 - 3 \mu m$, diameter 
$\sim 1 \mu m$ \cite{Grossman1982Changes}) compared to that of a Dd cell (length $\sim 22-25 \mu m$, diameter $\sim 4-6 \mu m$ \cite{barry2010dictyostelium,van2011amoeboid}),
the \textit{E.coli} can be well modeled as moving particles without considering the flow stirred by their deformation and movement.
Later (section \ref{Sec.MultiBac}) we will consider a system of one amoeboid Dd cell with a group of \textit{E. coli} in a dilute suspension, 
where the number of \textit{E. coli} is small ($\leq 12$) and are separated. In such a dilute suspension scenario, for simplicity, we do not consider the
hydrodynamic effects among the \textit{E. coli} or between the amoeboid Dd cell and a \textit{E. coli}. However,
we point out that if a large amount of \textit{E. coli} is presented, the active suspension will greatly alter the 
hydrodynamic effects of the system, causing effects including clustering of \textit{E. coli}. Refer to the Discussion section for 
potential future extensions.


We start with $N_B$ bacteria in the system, where each bacterium is represented as a dot with its position vector 
$\mathbf{x}_n = (x_n, y_n), \ n=1,2, \cdots, N_B$.
Without considering the flows generated by the movement of the bacteria, the movement of a bacterium mainly consists
of two parts: a convection term of the fluid, and an active movement term from the run-and-tumble. Since the mean
tumble interval ($\sim0.1$ sec) is much less than the mean run interval ($\sim1$ sec) \cite{Berg1990bacterial}, we model it as a random walk,
where the run is modeled as a jump and the tumble serves a reorientation of the bacterium.
The movement of each bacterium is described by the following equation:
\begin{eqnarray}\label{Eq.BacMotion}
 d \mathbf{x}_n  = \mathbf{u} ( \mathbf{x}_n) dt  + d \mathbf{X}_n
\end{eqnarray}
where $\mathbf{u}(\cdot)$ gives the fluid velocity field that is calculated via a complex analysis approach (section \ref{Sec.CellDef-Swim});
$  \mathbf{X}_n$ denotes the random walk of the bacterium, and we assume that at each small time step $d t$, the bacterium moves
a distance $\delta_J$ in the direction $\vartheta_n$. In the following discussion we start by considering a 2D random walk of the bacterium,
where $\vartheta_n \sim U(-\pi, \pi)$.

Recent research results reveal that bacterial rheotaxis plays a role in bacterial swimming, even without presence of
a nearby surface \cite{Fu2012Bacterial}.
To computationally investigate the effects of bacterial rheotaxis on bacterial escaping, that is, when the motions of the bacteria
are directed in response to the local fluid velocity gradient, we use a hybrid type of random walk model to model the bacterial rheotaxis, where 
the moving direction $\vartheta_n$ is given by the following equation:
\begin{eqnarray} \label{Eq.BRW-VarTheta}
\vartheta_n = (1-s_n)  \xi_r \pm  \big(  \arg {\mathbf{u} ( \mathbf{x}_n)} + s_n(1-s_n) \pi \xi_b \big)
\end{eqnarray}
where $s_n = \min (1, \|   \mathbf{u} ( \mathbf{x}_n   )   \|/M ) \in [0,1]$ measures the sensitivity of the bacterium to the local fluid velocity
with $M$ the cut-off value for the fluid velocity amplitude.
$\arg {\mathbf{u} ( \mathbf{x}_n)}$ is the argument of the local fluid velocity, $\xi_r  \sim U(-\pi, \pi) $ represents the random walk
part of $\vartheta_n$, $\xi_b \sim N(0, 1) $ and the sum of the two terms ($\arg {\mathbf{u} ( \mathbf{x}_n)} + s_n(1-s_n) \pi \xi_b$) represents the correlation with the fluid velocity due to
rheotaxis, where we are assume two types of rhoetactic movement -- along the flow ($+$) or against it ($-$). Equation (\ref{Eq.BRW-VarTheta})
is an empirical way to model the bacterial rheotaxis, in a way to ensure that 1) when the bacterium is far apart from the Dd cell ($ s_n \rightarrow 0$),
the bacterium does not sense the flow thus it undergoes a random walk without bias (notice that with $\xi_r  \sim U(-\pi, \pi) $, we have
$\xi_r \pm  \Theta    \sim U(-\pi, \pi) $ for any angle $\Theta$), 2) when the bacterium is near the Dd cell ($s_n \rightarrow 1$),
bacterial movement is dominated by rhotaxis ($\vartheta_n = \pm  \arg {\mathbf{u} ( \mathbf{x}_n)}$), and 3) bacterial movement
continuously change between unbiased and biased random walk depending on $s_n$.

Comparing to a typical shape deformation cycle of a Dd cell of about $T \sim 1-2$ min \cite{barry2010dictyostelium,van2011amoeboid}, 
the mean run interval is $\sim 1$ sec subjected to exponential distribution \cite{Berg1990bacterial}. For simplicity, we take the small time step $dt$
of the random walk $d \mathbf{X}_n$ as $dt = 0.1 T$, where $T$ is the average period of a Dd cell swimming cycle.


\subsection{Chemotaxis signaling dynamics.}\label{Sec.SignDyn}

\textit{Dictyostelium discoideum} (Dd) utilizes folic acid receptor 1 (fAR1), a class of single G-protein-coupled receptor (GPCR) 
to detect diffusible chemoattractant folate secreted by bacteria, thus to locate and chase bacteria \cite{pan2018g}.
Once the amoeba ``catches" the bacteria, the amoeba engulfs and consumes them. Dd amoeboid cell is reported
to ingest, kill and digest bacteria at a rate of at least one per minute \cite{cosson2008eat}. 

Suppose that at time $t$, the amoeboid cell captures a region $\Omega_{\textrm{Dd}} (t)$ in the 2D infinite domain, thus 
$\Omega_{\textrm{Dd}}^C, \ \partial \Omega_{\textrm{Dd}} $ give the
external fluid domain and the cell boundary, respectively.
Let $f (\mathbf{x}, t)$, $R_f^0 (\mathbf{x},t)$ and $R_f (\mathbf{x},t)$ denote the concentration of diffusive folate,
the surface concentration of free fAR1 receptors and the surface concentration of the
folate-bond fAR1 receptors, respectively. $f (\mathbf{x}, t)$ is defined on $\Omega_{\textrm{Dd}}^C (t) \times [0, \infty)$ and
$R_f^0 (\mathbf{x},t)$, $R_f^0 (\mathbf{x},t)$ are defined on $ \partial \Omega_{\textrm{Dd}} (t) \times [0, \infty) $.
The signaling dynamics of the diffusible chemoattractant folate and the fAR1 receptors on the cell membrane are modeled by the following 
reaction-diffusion-convection (RDC) equations together with boundary conditions:
\begin{eqnarray}\label{Eq.fol}
\frac{\partial f }{\partial t} &=& D \Delta f - \mathbf{u}   \cdot \nabla f  
 +  a \int \sum_{n=1}^{N_B}   \delta (\mathbf{x} - \mathbf{x}_n) d \mathbf{x}    \qquad \textrm{in} \  \Omega_{\textrm{Dd}}^C (t) \\  \label{Eq.fol_bd_cell}
 D \frac{\partial f}{\partial n} &=&   - k_+ f R_f^0 + k_- R_f    \qquad \textrm{on} \  \partial \Omega_{\textrm{Dd}}  (t)  \\        \label{Eq.fAR}
\frac{\partial R_f }{\partial t} &=&  k_+ f R_f^0 - k_- R_f   - \gamma R_f  + \varsigma R_f \frac{dW}{dt}   \qquad \textrm{on} \  \partial \Omega_{\textrm{Dd}}  (t)
\end{eqnarray}
with the constraints:
$$ R_f^0 (\mathbf{x},t) + R_f (\mathbf{x},t) = R_{\max} , \qquad  f (\mathbf{x}, t) \geq 0, \  R_f^0 (\mathbf{x},t) \geq 0, \ R_f  (\mathbf{x},t) \geq 0$$
The terms in equations (\ref{Eq.fol} - \ref{Eq.fAR}) are explained as follows.
\begin{itemize}
\item $D \Delta f $: diffusion of folate with the diffusion rate $D$. 
\item $\mathbf{u}   \cdot \nabla f $: convection of folate, where $\mathbf{u}$ gives the velocity field of the extra-cellular flow.
\item $ a \int \sum   \delta  d \mathbf{x}  $: production of folate molecules from the bacteria, where $a $ is the folate production
rate. For simplicity, we assume that all bacteria have the same folate production rate.
\item $ k_+ f R_f^0, \ k_- R_f  $: 
biochemical reactions between folate molecules and fAR1 receptors along the cell membrane boundary, where $k_+, \ k_-$ 
are the binding and unbinding rates of the fAR1 receptors.
$R_{\max}$ is the sum of free and folate-bond 
receptors, and we assume it a constant along the cell boundary.  
\item $ \gamma R_f  $: degradation of folate-bond fAR1 receptors, where $\gamma $ is the degradation rate.
\item $ \varsigma R_f dW$: white noise that captures stochastic effects in intracellular signal dynamics, where $\varsigma$
is the noise strength.
 
\end{itemize}

\begin{figure}[h]
\center
  \includegraphics[width=0.9\textwidth]{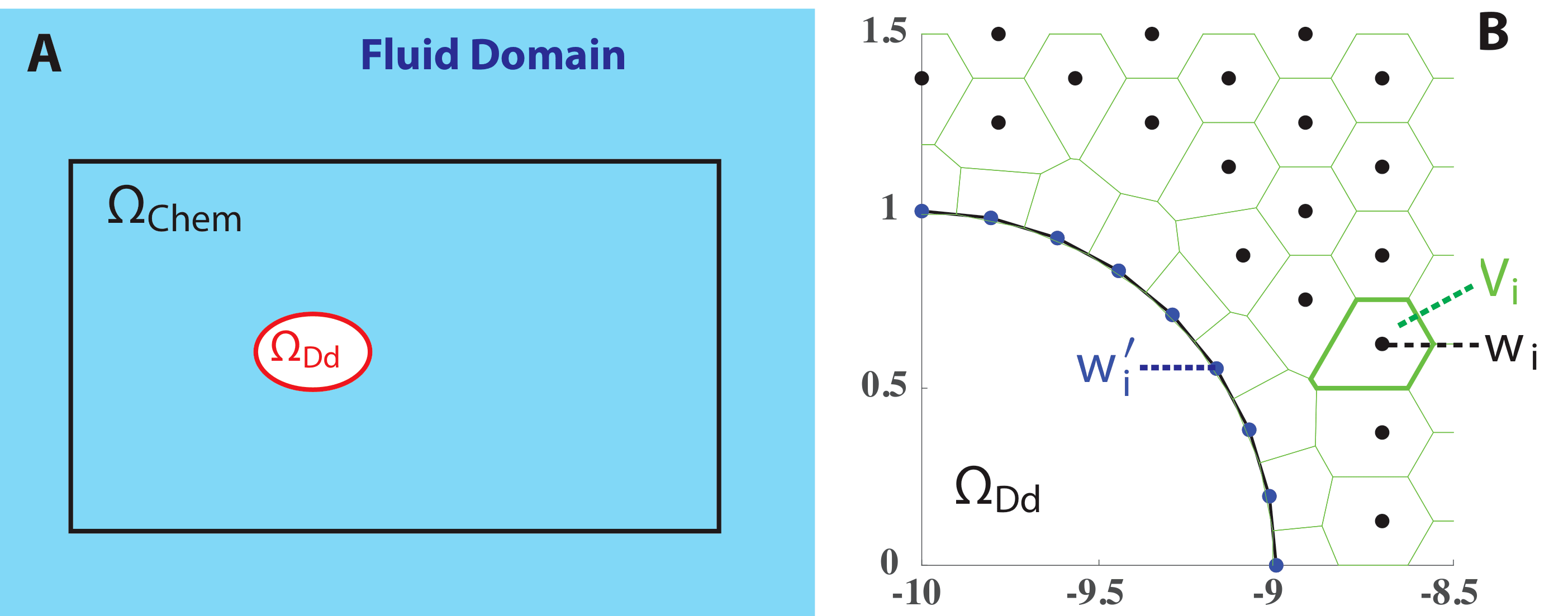}
  \caption{\textbf{A} Illustration of the geometry of the computational domains. \textbf{B} A local view of the Voronoi meshes
  near the Dd cell. }
  \label{Fig.NumGeo}
\end{figure}

Computationally, instead of considering the infinite fluid domain, we consider a finite but large enough computational domain $\Omega_{\textrm{Chem}}$
that contains the cell domain $\Omega_{\textrm{Dd}}$ and all the bacteria, and $\textrm{Area} ( \Omega_{\textrm{Dd}} ) \ll \textrm{Area} ( \Omega_{\textrm{Chem}} )$
(illustrated in figure \ref{Fig.NumGeo}A). Therefore 
the fluid domain boundary consists of two pieces: $\partial \Omega_{\textrm{Dd}}$ and $\partial \Omega_{\textrm{Chem}}$. We assume no-flux Neumann
boundary condition $\hat{\mathbf{n}} \cdot \nabla f  =0$ on $\partial \Omega_{\textrm{Chem}}$.

To solve the RDC equations (\ref{Eq.fol}-\ref{Eq.fAR}), we use the Voronoi tessellation based finite volume method formulated in
\cite{bottino2001computer,dillon2008single}. 
Initially, we generate a network of fluid ``nodes" $\{ \mathbf{w}_i \}$ 
in the computational fluid domain $\Omega_{\textrm{Chem}} \cap \Omega_{\textrm{Dd}}^C$,
and discretize the cell boundary $\partial \Omega_{\textrm{Dd}}$ by $N_R$ nodes $\mathbf{w}'_0,  \mathbf{w}'_1, 
\cdots,$ $ \mathbf{w}'_{N_R} = \mathbf{w}'_0$ - how to choose the discretization will be discussed in section \ref{Sec.CellDef-Swim}.
Notice that the positions of both the boundary nodes $\{ \mathbf{w}_i' (t) \}$ and $\{ \mathbf{w}_i (t) \}$ change as the Dd cell deforms
and perturbs the surrounding fluid. At each computational time step,  the fluid nodes $\{ \mathbf{w}_i \}$ are updated as 
\begin{eqnarray}\label{Eq.UpdateMesh}
\mathbf{w}_i^{n+1} = \mathbf{w}_i^n + \mathbf{u} (\mathbf{w}_i^n) \Delta t 
\end{eqnarray}
where $\mathbf{u} $ is the fluid velocity field.
We generate a Voronoi tessellation $\{ V_i^{n+1} \} \cup \{ {V_i'}^{n+1} \} $ of the 
computational domain $\Omega_{\textrm{Chem}} \cap \Omega_{\textrm{Dd}}^C$ based on the nodes $\{ \mathbf{w}_i^{n+1} \} \cup \{ {\mathbf{w}_i'}^{n+1} \}$,
that is, $V_i^{n+1}   $ (or ${V_i'}^{n+1}$) is the set of all the points in the computational fluid domain $\Omega_{\textrm{Chem}} \cap \Omega_{\textrm{Dd}}^C$ 
closer to the node $\mathbf{w}_i$ (or $\mathbf{w}_i'$) than any other node (figure \ref{Fig.NumGeo}B).
 
The folate concentration data $f$ is available at
the nodes $\{ \mathbf{w}_i \}$. In the Lagrangian frame, the convection term $\mathbf{u} \cdot\nabla f$ disappears from
equation (\ref{Eq.fol}). 
The Laplacian in equation (\ref{Eq.fol}) can thus be approximated for each Voronoi tile $V_i$ through summation
of the fluxes across the edges partitioning $V_i$ from each of its Delaunay neighbors (denoted by $\Lambda_i$) \cite{bottino2001computer}:
\begin{eqnarray}\label{Eq.NumLap}
\Delta f_i \approx  \frac{1}{\textrm{Area} (V_i)} \int_{\partial V_i} \mathbf{n} \cdot \nabla f   d s  
\approx  \frac{1}{\textrm{Area} (V_i)} \sum_{j \in \Lambda_i} \frac{f_j - f_i}{\| \mathbf{w}_j - \mathbf{w}_i \|} l_{ij}  
\end{eqnarray} 
where $f_i = f(\mathbf{w}_i)$, and $l_{ij}$ is the length of the common edge shared by $V_i$ and $V_j$. 
Notice that for a Voronoi tile $V_i$, its Delaunay neighbors may also include Dd cell boundary tiles $V_i'$, 
but for simplicity, we omit the notation ' in equation (\ref{Eq.NumLap}). 
Numerical convergence studies show that the method converges linearly in the $L^2$ norm \cite{borgers1987lagrangian}.
We solve equations (\ref{Eq.fol}) numerically in a forward Euler scheme, with 
the Laplacian approximated by equation (\ref{Eq.NumLap}). 
Equation (\ref{Eq.fol}) can be numerically solved as follows:
\begin{eqnarray*}
f_{i}^{n+1} &=& f_{i}^{n} + \left( D  \Delta f_i^n + \frac{ a \sum_{n=1}^{N_B}  \delta_{V_i} (\mathbf{x}_n) }{\textrm{Area} (V_i)} \right) \Delta t   
\end{eqnarray*}
where $\delta_{V_i} (\mathbf{x}_n) = 1 $ if the $n$th bacteria is in $V_i$, otherwise $\delta_{V_i} (\mathbf{x}_n) = 0$.

For boundary conditions, first we notice that while other nodes $\mathbf{w}_i$ locates inside the corresponding Voronoi tile $V_i$, the cell boundary nodes $\mathbf{w}_i'$
locate on $\partial V_i' \cup \partial \Omega_{\textrm{Dd}}$ (figure \ref{Fig.NumGeo}B). 
To each cell boundary $\mathbf{w}_i'$, the numerical Laplacian equation (\ref{Eq.NumLap}) can be modified as:
\begin{eqnarray} \label{Eq.NumLap_BDnode}
\Delta f_i \approx  \frac{1}{\textrm{Area} (V_i')} \Big( \sum_{j \in \Lambda_i} \frac{f_j - f_i}{\| \mathbf{w}_j - \mathbf{w}_i' \|} l_{ij}  
+ \big( \mathbf{n} \cdot \nabla f \big) l_i'    \Big) 
\end{eqnarray}
where we approximate the boundary
$\partial V_i \cap  \partial \Omega_{\textrm{Dd}} $ by a line segment connecting the two vertices shared by neighboring Voronoi tiles, and let
$l_i'$ be its length.
For boundary condition on $ \partial \Omega_{\textrm{Dd}} $ given by equation (\ref{Eq.fol_bd_cell}), along the small boundary segment $\partial V_i' \cap \partial \Omega_{\textrm{Dd}} $ we have:
\begin{eqnarray*}
\frac{\partial f}{\partial n} = \mathbf{n} \cdot \nabla f =  \frac{1}{D} \big( - k_+ f R_f^0 + k_- R_f  \big) 
\end{eqnarray*}
and equation (\ref{Eq.NumLap_BDnode}) becomes:
\begin{eqnarray*}
\Delta f_i \approx  \frac{1}{\textrm{Area} (V_i)} \Big( \sum_{j \in \Lambda_i} \frac{f_j - f_i}{\| \mathbf{w}_j - \mathbf{w}_i' \|} l_{ij}  
+ \frac{l_i^M }{D} \big( -k_+ f R_f^0 + k_- R_f \big)     \Big)  
\end{eqnarray*}
Finally, the no-flux boundary condition on $\partial \Omega_{\textrm{Chem}}$ can be directly enforced to equation (\ref{Eq.NumLap_BDnode}).


\subsection{Chemotaxis induced Dd shape deformations and swimming.}\label{Sec.CellDef-Swim}

When adhesion is absent thus cell crawling is disabled, Dd cells can swim towards a chemoattractant source.
During swimming, cells form pseudopods that convert to rear-ward moving bumps thereby propelling itself through the 
surrounding fluid in a totally adhesion-free fashion
\cite{bae2010swimming,barry2010dictyostelium,van2011amoeboid,howe2013amoebae}. 
Such a swimming mode
is very different from ciliated or flagellated swimming modes that are commonly adopted by many bacteria including
\textit{E. coli}, as it is the one that requires large deformations that propagate over the cell body. 
We use the \textit{mathematical amoeba model} 
\cite{shapere1987self,shapere1989geometry,avron2004optimal,Bouffanais2010hydrodynamics,wang2016computational} 
to generate the Dd cell deformation as well as to solve the resulting cell-fluid interaction. 
In the following we list the outline of the modeling framework, see the Appendix for more details.

Consider the following conformal mapping defined from $\{\zeta \in$ \ifiopams$\mathbb{C}$\fi; $ | \zeta | \geq 1 \}$ to
$\Omega_{\textrm{Dd}}^C$:
\begin{eqnarray}\label{Eq.AmoebaConfMap-2Term}
w(\zeta;t) = e^{i \theta (t)} \Big[ r(t)  \zeta  + \frac{\eta_{-1} (t)}{\zeta} + \frac{\eta_{-2} (t)}{\zeta^2} \Big] + Z_{\textrm{Dd}}(t)
\end{eqnarray}
where the Dd cell shape is defined by $\partial \Omega_{\textrm{Dd}} (t) = \{ w (\sigma ;t) | \sigma \in S^1\}$. 
The $N_r$ discretization nodes $ \{ \mathbf{w}'_1, \mathbf{w}'_2, \cdots, \mathbf{w}'_{N_r} \}$ are generated as follows: 
we discretize the unit circle $\partial D$ in the computational $\zeta$-plane equally into $N_r$ nodes:
\begin{eqnarray*}
\sigma_j = e^{i\theta_j} = e^{i\frac{2 \pi j}{N_r}}, \qquad j=0, 1,2,\cdots, N_r-1
\end{eqnarray*} 
then let $ \mathbf{w}'_j = w(\sigma_j;t)$.
In Eq (\ref{Eq.AmoebaConfMap-2Term}),
$ \theta (t) \in \ifiopams\mathbb{R}\fi$ gives the cell polarization that will be determined by signaling sensing dynamics as 
discussed below. The swimming Dd cell undergoes cyclic shape deformation with the same period $T$. 
We assume that the polarization $\theta$ is determined at the beginning of
a swimming cycle and will not change during the cycle, thus $\theta (nT + t) \equiv \theta (nT)$ for $t \in[0,T)$. 
$r , \eta_{-1}, \eta_{-2} \in \ifiopams\mathbb{R}\fi$ control the cell size and shape deformations, and are subjected
to area conservation of the cell. $Z_{\textrm{Dd}}(t)$ gives the location of the cell, while $U_{\textrm{Dd}}(t) = \dot{Z}_{\textrm{Dd}}(t)$ 
gives the velocity of the cell and is computed from the Goursat formula \cite{muskhelishvili2013some}. The fluid
velocity field $\mathbf{u}$, or in the complex notation $u$, can be also obtained through the Goursat formula. 
Refer to \ref{Sec.2DLRNSwim} for more details of the complex analysis techniques involved in this part.

\begin{figure}[h]
\center
  \includegraphics[width=0.9\textwidth]{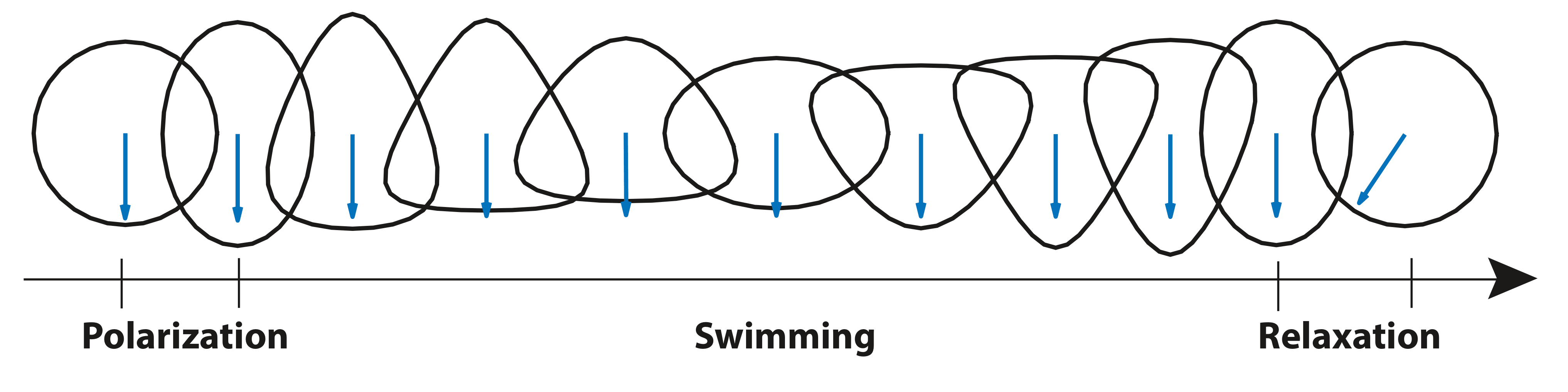}
  \caption{Shapes of the Dd cell in a cycle, showing the polarization, swimming and relaxation phases. 
  Arrows indicate the cellular polarization direction $\theta$ within the cycle. At the end of the cycle, a new direction is selected. }
  \label{Fig.CycleShape}
\end{figure}

We assume that the Dd cell undergoes shape deformations in response to signal gradient, with
each swimming cycle lasting for a period of $T$ and consisting of three phases: (i) \textit{polarization}, 
when the Dd cell determines the polarization $\theta$ during the current cycle in response to the signal $R_f$,
and elongates its cell body in preparation for (ii) \textit{swimming}, when the Dd cell deforms its shape so to
actively swim along the polarization direction, followed (iii) \textit{relaxation}, when the Dd cell returns to its initial
circular shape. Durations of polarization and relaxation phases are chosen to be much shorter than the swimming phase.
Figure \ref{Fig.CycleShape} shows a typical cycle of the Dd cell shape deformations. For more details of the modeling setup 
of the signaling induced Dd cell polarization and shape deformations, please refer to \ref{Sec.Pol-Def}.

In our model, we do not model the engulfment process in phagocytosis. 
For simplicity, once a bacterium falls within a close enough distance near a cell boundary node $\mathbf{w}_j'$,
we consider it taken by the amoeba and remove it from the system.

\subsection{Computations of the model}

We nondimensionalize the system, using the duration of a Dd cell swimming cycle $T$ and the radius of a Dd cell at rest $r_0$
as the characteristic temporal and spatial scales, and $R_{\max} $ the characteristic concentration scale for $f$ and $R_f$. The 
non-dimensionalized system can be found in \ref{Sec.Non-Dim-Supp}.

The system is computed using the following update algorithm: 

\begin{enumerate} 
\item \textbf{Signaling dynamics.} Generate the Voronoi tessellation from the current distribution of ``nodes" $\{\mathbf{w}_i  \} \cup \{ \mathbf{w}_i' \}$.
Update $f$ and $R_f $ by solving the RDC equations using the finite volume method in the moving mesh.
\item \textbf{Dd amoeboid cell shape.} 
\begin{itemize} 
\item \textbf{If} the cell is at rest in a circular shape ($t = nT, \ n \in \ifiopams \mathbb{N} \fi  $), determine the cell polarization $\theta $ for this swimming cycle.
\item \textbf{ElseIf} the cell is during a swimming cycle, $t \in (nT, (n+1)T), \ n \in \ifiopams\mathbb{N} \fi   \}$, update the conformal mapping $w$ equation (\ref{Eq.AmoebaConfMap-2Term}).
\end{itemize}
\item \textbf{Fluid mechanics.} Update flow velocity field $\mathbf{u}$ from the Goursat formulas. Update
bacteria positions and the moving mesh:
\begin{itemize}
\item \textbf{Bacteria motions.} Update bacterial positions by equation (\ref{Eq.BacMotion}). Remove any bacterium that 
comes to cell boundary Voronoi tiles.
\item \textbf{Moving mesh.} Update the moving mesh nodes $\{\mathbf{w}_i  \}$ from equation (\ref{Eq.UpdateMesh}).
\end{itemize}
\end{enumerate}

The numerical scheme works for our model system where hydrodynamics and signaling dynamics are coupled, and 
allows us to study cellular chemotaxis and rheotaxis in a fluid environment. We point out that we do see fluid nodes
near the Dd cell boundary getting closer after many swimming cycles due to the large deformation of the cell, which causes numerical instability of
the finite volume method. We solve this numerical issue by a local re-mesh near the Dd cell, see more details in \ref{Sec.Remesh-Supp}. Parameter values used in our simulations are given in Table \ref{Tab.Para}.


\section{Results}\label{Sec.Results}

\subsection{Chemotaxis guided amoeboid swimming allows the Dd amoeboid cell to follow and catch bacteria} \label{Sec.ChemoDd}

\begin{figure}[h]
\center
  \includegraphics[width=0.95\textwidth]{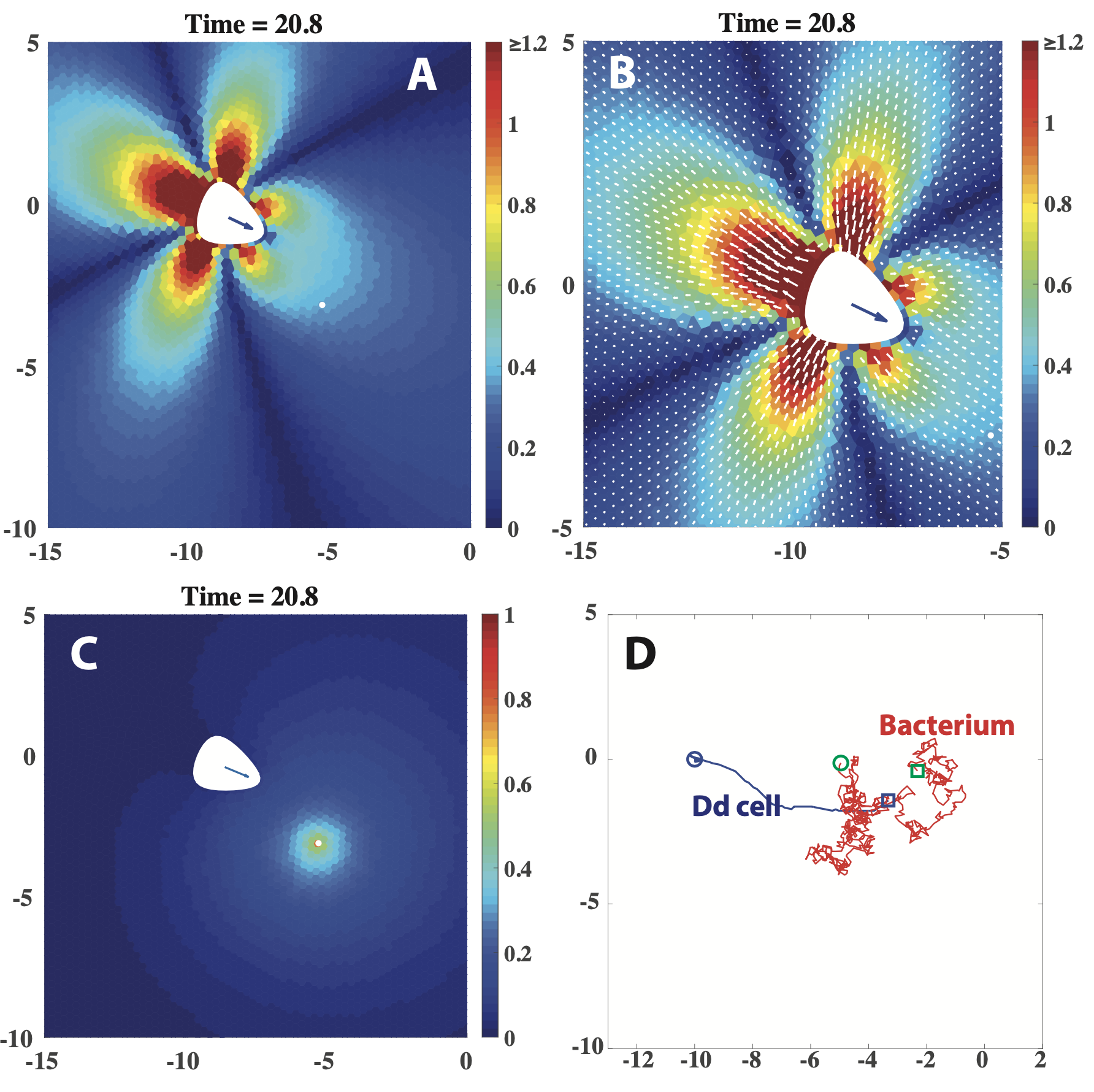}
  \caption{\textbf{Simulation of a swimming Dd cell following a bacterium guided by chemotaxis. AB} 
Snapshots of the fluid velocity at $T=20.8$, where the heatmap (\textbf{B}) shows the amplitude of the fluid velocity,
in the zoom in view near the Dd cell (\textbf{C}),  the white arrows show the fluid velocity directions. \textbf{C}
A snapshot of the heatmap of the diffusive folate signal concentration at $T=20.8$. In \textbf{ABC}, both the Dd cell
  and the bacterium are colored in white. \textbf{D} trajectories of the Dd cell (blue) and the bacterium (red), where circles
  show the start points of the Dd cell center (blue circle) and the bacterium (green circle), and squares show the end points
  of the Dd cell center (blue square) and the bacterium (green square). The Dd cell catches the bacterium after 69 swimming cycles.}
  \label{Fig.BaUNqZ68}
\end{figure}

We start with a system consisting of one Dd cell and one bacterium where the bacterium undergoes an unbiased random walk (equation (\ref{Eq.BacMotion})).
Simulation results show that the Dd amoeboid cell is able to swim following the bacterium guided by the chemoattractant signal. Figure \ref{Fig.BaUNqZ68}
shows a typical simulation, where the Dd cell catches the bacterium after 69 cycles. Compared to the random walk of the bacterium, 
the motion of the Dd cell is more directed; as the Dd cell approaches 
the bacterium, the bacterium is pushed away to the right by the flow generated by the swimming Dd cell,
and eventually being captured by the Dd cell (figure \ref{Fig.BaUNqZ68}C). The full time lapse snapshots of the fluid velocity 
profile during one Dd cell swimming cycle is provided in Figure \ref{Fig.BaUNqZ68-Cycle}.

\begin{figure}[h]
\center
\includegraphics[width=0.95\textwidth]{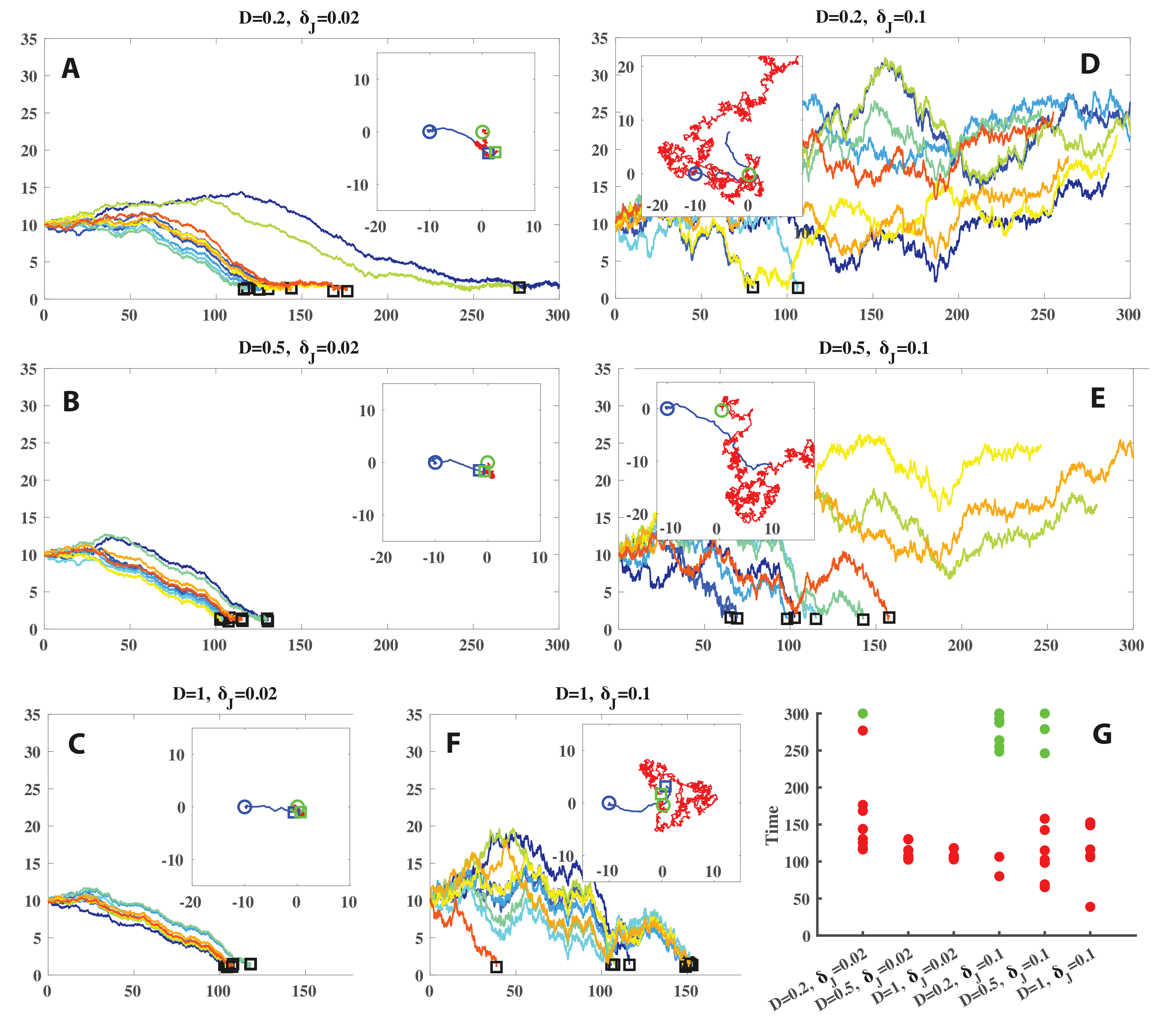}
\caption{ \textbf{Simulation results of chemotaxis guided Dd cell swimming. A-F} Distance between the Dd cell center
and the bacterium with different combinations of $D$ and $\delta_J$ values. $x$-axis shows the time counted as 
swimming cycles. Each colored curve shows the result from one simulation from the group, and black 
boxes show when the Dd cell catches the bacterium.
The sub-panel in each panel shows the trajectories of the Dd cell (blue curve) and the bacterium (red curve) 
from one typical simulation from the group, with the blue and green circles mark the starting locations
  of the Dd cell and the bacterium, and the blue and green squares mark the end locations
  of the Dd cell and the bacterium when the Dd cell catches the bacterium.
\textbf{G} Durations of the simulations, where red dots show when the Dd cell catches
the bacterium, green dots show when the bacterium runs out of $\Omega_{\textrm{Chem}}: [-25, 25] \times [-25, 25]$, or the Dd cell
does not catch the bacterium by the end of 300 cycles.  }
  \label{Fig.CellBac_DJ}
\end{figure}

We consider how the chemoattactant diffusion rate $D$ and the bacterial jump amplitude $\delta_J$ affect the 
Dd cell swimming dynamics. We perform 6 groups of 10 simulations, with different values of $D$ and $\delta_J$: 
$D=0.2, \ 0.5, \ 1, \ \delta_J =0.02, \ 0.1$. Simulations results show large coefficient rate $D$ makes it easier for
the Dd cell to follow the signal guidance thus catch the bacterium; bacterial random walk strength $\delta_J$
may also help the bacterium to escape, however, the more important effect of $\delta_J$ is that it increases
the variance of time for the Dd cell to catch the bacterium, if it could (figure \ref{Fig.CellBac_DJ}). The 
increase of catch time variance needed caused by $\delta_J$ should not be surprising, since the passive 
motion of the bacterium is an unbiased random walk $\mathbf{X}$, thus $E(\mathbf{X})=0$ and large $\delta_J$
only increases $\textrm{Var} (\mathbf{X})$ which in return increases the catch time variance, though through
a complex chemotaxis induced amoeboid swimming dynamics. Figure \ref{Fig.CellBac_DJ} shows that
in 10 out of 10 simulations with $D=0.5, \ \delta_J = 0.02$ (figure \ref{Fig.CellBac_DJ}C) $D=1, \ \delta_J = 0.02$ (figure \ref{Fig.CellBac_DJ}C)
or $D=1, \ \delta_J = 0.1$ (figure \ref{Fig.CellBac_DJ}F), the Dd cell is able to catch the bacterium within 300 cycles,
compared to 9 out of 10 with $D=0.2, \ \delta_J = 0.02$ (figure \ref{Fig.CellBac_DJ}A) and
7 out of 10 with $D=0.5, \ \delta_J = 0.1$ (figure \ref{Fig.CellBac_DJ}E). The worst
scenario for the Dd cell is small $D$ and large $\delta_J$: with $D=0.2, \ \delta_J = 0.1$,
only in 2 out of 10 simulations the Dd cell catches the bacterium within 300 cycles (figure \ref{Fig.CellBac_DJ}D).
In the above simulations we take $k_- = 0$, finally we take $k_-= 0.1$ and with $D = 0.5, \ \delta_J = 0.02$,
the results from 10 simulations are shown in Figure \ref{Fig.CellDist_Km}, compare with $k_-=0$ (figure \ref{Fig.CellBac_DJ}B),
which indicates that large $k_-$ helps the bacterium to escape from the Dd cell.

We present the time lapse snapshots from typical simulations with $\delta_J = 0.1, \ D = 0.2, \ 0.5 , \ 1$
in Figures \ref{Fig.Mcf7QdMF} - \ref{Fig.HuyRtX3j}.

\subsection{Rheotaxis helps bacteria escaping from the predator} \label{Sec.RheoBac}

Next, we use the model to investigate if rheotaxis could help the bacterium run away from the predator Dd cell. In particular,
we consider two types of bacterial rheotaxis: the bacterium prefers to move with the flow vs. against the flow. We model either of the two
rheotactic systems by a biased random walk of the bacterium, where the direction of the bacterium's jump $\vartheta$ is given by equation
(\ref{Eq.BRW-VarTheta}), and it takes the $+$ sign if the bacterium prefers to move with the flow, and the $-$ sign if against the flow.

 \begin{figure}[h]
\center
  \includegraphics[width=0.95\textwidth]{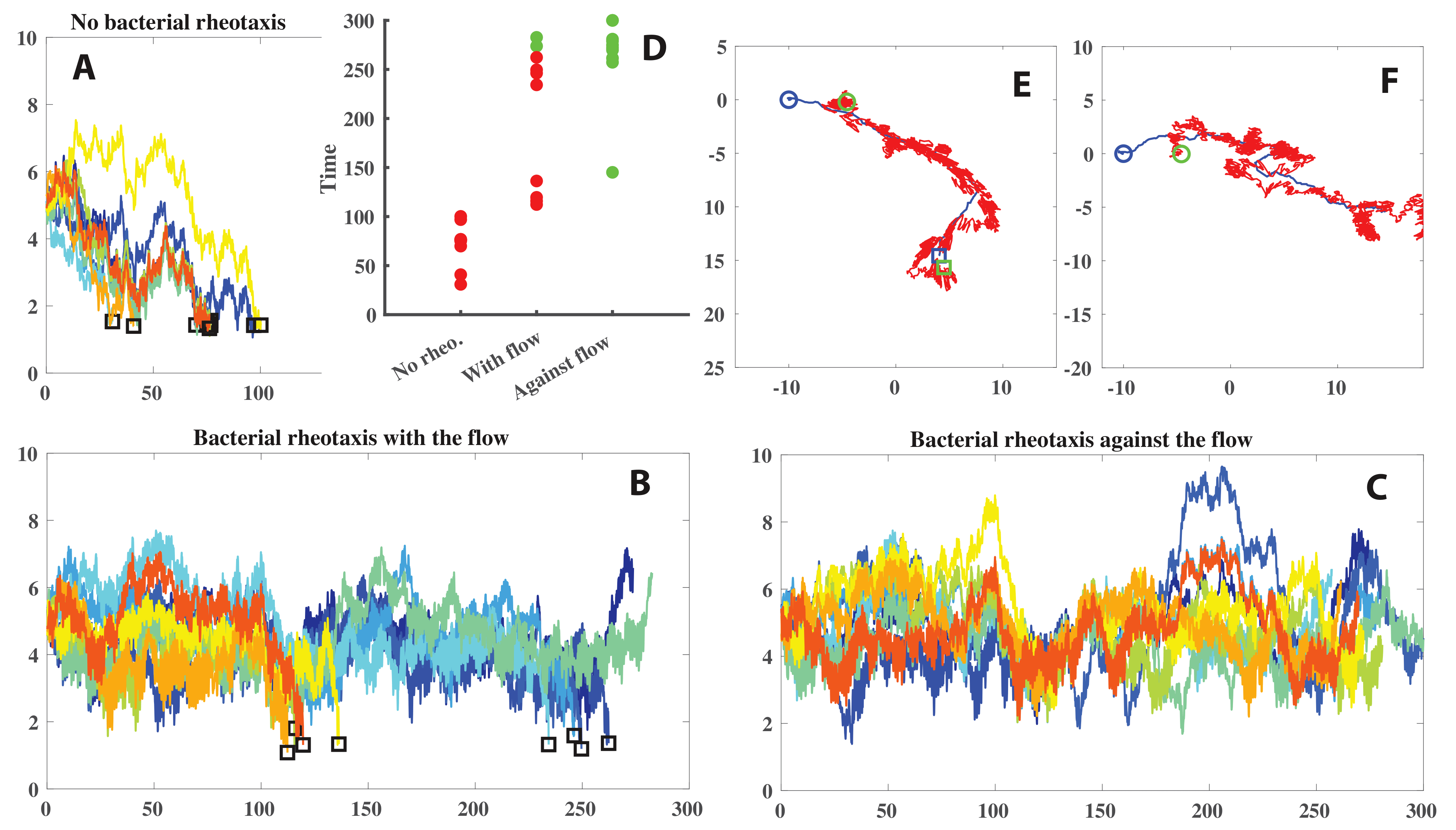}
  \caption{\textbf{Simulation results of bacterial rheotaxis. ABC} Distance between the bacterium and the Dd cell
  center with \textbf{A} no bacterial rheotaxis, \textbf{B} bacterial rheotaxis with the flow, and \textbf{C} bacterial 
  rheotaxis against the flow. $x$-axis shows the time counted as 
swimming cycles. Each colored curve shows the result from one simulation from the group, and black 
  boxes in \textbf{ABC} show when the Dd cell catches the bacterium.
  \textbf{D} Durations of the simulations, where red dots mark the time when the Dd cell catches
  the bacterium, green dots mark when the bacterium runs out of $\Omega_{\textrm{Chem}}: [-25, 25] \times [-25, 25]$, or the Dd cell
  does not catch the bacterium by the end of 300 cycles. \textbf{E} Trajectories of the Dd cell (blue curve) and the 
  bacterium (red curve) with bacterial rheotaxis with the flow, with the blue and green circles mark the starting locations
  of the Dd cell and the bacterium, and the blue and green squares mark the end locations
  of the Dd cell and the bacterium when the Dd cell catches the bacterium. \textbf{F} Trajectories of the Dd cell (blue curve) and the 
  bacterium (red curve) with bacterial rheotaxis against the flow, with the blue and green circles mark the starting locations
  of the Dd cell and the bacterium, the Dd cell does not catch the bacterium in 300 cycles.}
  \label{Fig.BacRheo}
\end{figure}

We performed three groups of simulations: 1) no bacterial rheotaxis, 2) bacterial rheotaxis with a biased random walk with the flow,
and 3) bacterial rheotaxis with a biased random walk against the flow. We take $D = 1, \ \delta_J = 0.05$ in all three groups, and
in each group we run 10 simulations. Simulation results are shown in figure \ref{Fig.BacRheo}. Without bacterial rheotaxis,
the Dd cell catches the bacterium in all 10 simulations (figure \ref{Fig.BacRheo}AD). With bacterial rheotaxis with the flow, in 8
out of 10 simulations it needs longer time for the Dd cell to catch the bacterium and 8 out of 10, and in the other 2 simulations
the bacterium runs out of the $\Omega_{\textrm{Chem}}$ domain after $>250$ Dd cell swimming cycles (figure \ref{Fig.BacRheo}BD).
Finally with bacterial rheotaxis against the flow,  in 8 out of 10 simulation the bacterium runs out of the $\Omega_{\textrm{Chem}}$ domain,
and in the other 2 simulations the bacterium stays within the $\Omega_{\textrm{Chem}}$ domain always but the Dd cell
does not catch the bacterium in $300$ cycles (figure \ref{Fig.BacRheo}CD). Figure \ref{Fig.BacRheo}E shows the trajectories of the Dd cell and the 
bacterium with bacterial rheotaxis with the flow from one simulation, when the Dd cell catches the bacterium at the end, and 
figure \ref{Fig.BacRheo}F shows the trajectories of the Dd cell and the 
bacterium with bacterial rheotaxis against the flow from another simulation, the Dd cell does not catch the bacterium in 300 cycles.

Our simulation results (figure \ref{Fig.BacRheo}) show that while bacterial rheotaxis against the flow appears to be a good strategy
for the bacterium to escape from the predator Dd cell, bacterial rheotaxis with the flow may also help the bacterial to run away
a little compared to no rheotaxis. We emphasize that at LRN, since inertia is absent, the current flow profile only depends on
the current Dd cell deformation, and it keeps changing over a swimming cycle (see Figure \ref{Fig.BaUNqZ68-Cycle}). The bacterial rheotaxis we find here
is the result out of one or even multiple Dd cell swimming cycles. Moreover,
we would also point out that even in the best simulated scenario -- bacterial
rheotaxis against the flow, the bacterium is not able to fully escape from the Dd cell, which can be seen from figure \ref{Fig.BacRheo}C
that the distance between the Dd cell and the bacterium oscillates but the mean is not increasing, indicating that the Dd cell
keeps following the bacterium.


\subsection{Chemotaxis guided amoeboid swimming caused by a dilute suspension of bacteria} \label{Sec.MultiBac}

Finally we consider the system with one Dd cell and a dilute suspension of bacteria, that is, a small number of bacteria present in the system.
In the dilute suspension, the hydrodynamic interactions among the bacteria can be ignored. 
We consider systems where initially a group of $N$ bacteria locate equispaced in a ring with the Dd cell center
as the ring center (see supplement movie and Figures \ref{Fig.mq7Ku6JH} - \ref{Fig.ssI3XJjA}, the color scales are the same as
figure \ref{Fig.BaUNqZ68}C), and $D=1, \ \delta_J = 0.1$ in all simulations in this section.

Simulations results show that similar to the systems with only one bacterium, when multiple bacteria are present, the Dd cell
is still able to follow the chemoattractant signals and catch some bacteria. However, we notice that unlike in the systems with only
one bacterium where the Dd cell movement is more directed, now with more bacteria present, the movement of the Dd cell 
becomes more chaotic -- the Dd cell changes its polarization between consecutive cycles more frequently. To quantify this effect,
we define a Dd cell polarized orientation change index PCI as follows:
\begin{eqnarray*}
\textrm{PCI}_n = \frac{1 - \cos \big( \theta_{n} - \theta_{n-1} \big)}{2}, \qquad n \geq 2, \ n \in    \ifiopams\mathbb{N}\fi
\end{eqnarray*} 
where $\textrm{PCI}_n$ denotes the PCI for the $n$th swimming cycle $t \in [nT, (n+1) T)$, and $\theta_n $ is the Dd cell polarization
angle in the $n$th cycle. $\textrm{PCI}_n$ takes values in $[0,1]$, with $\textrm{PCI}_n=0$ means that the polarization does not 
change between the two consecutive cycles, while $\textrm{PCI}_n=0$ means that the polarization changes with an angle $\pi$,
i.e., the polarization is reversed.

\begin{figure}[h]
\center
  \includegraphics[width=0.95\textwidth]{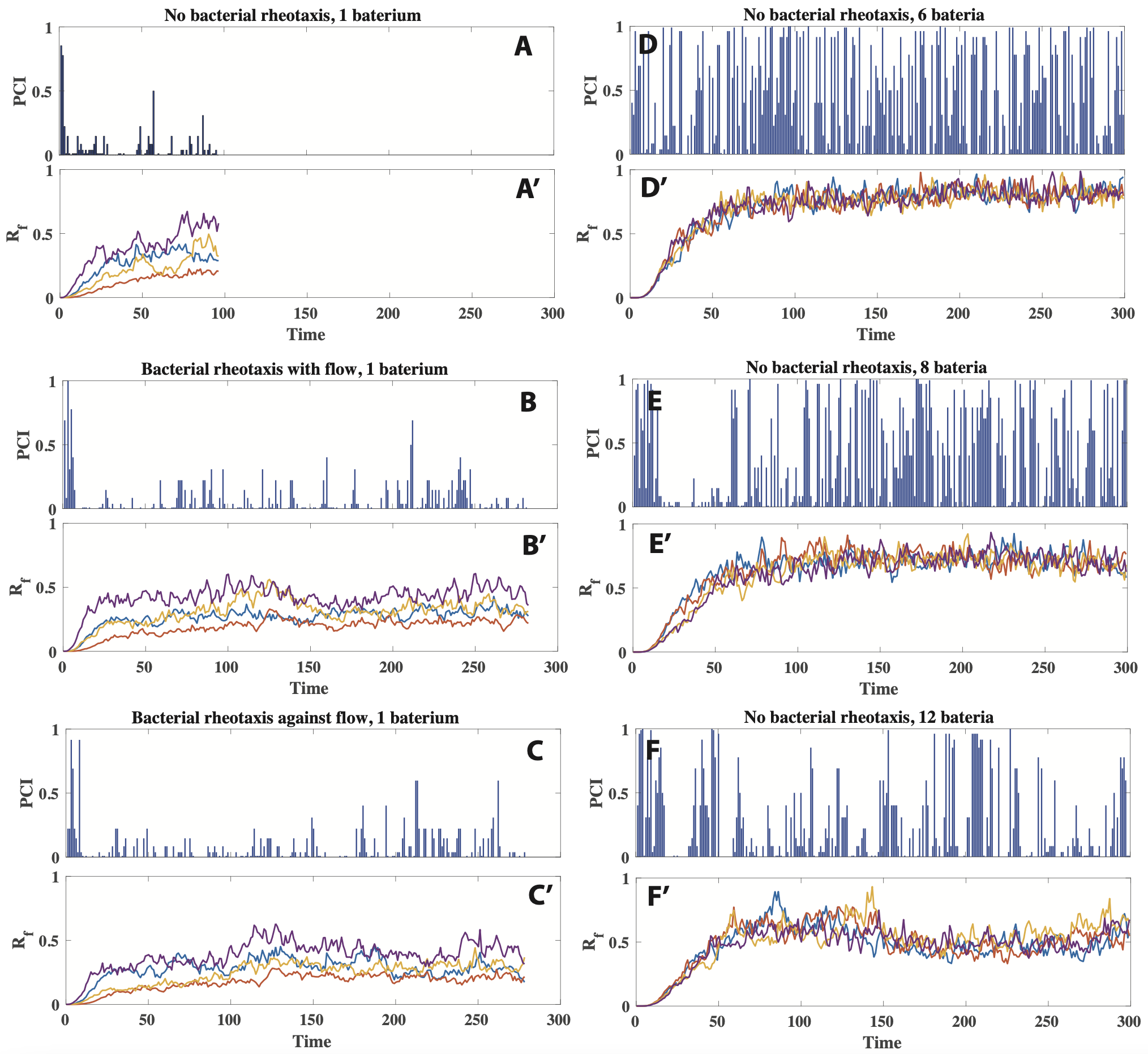}
  \caption{\textbf{Dd cell polarization change.} Each top panel (\textbf{A} - \textbf{F}) shows the PCI changes in one simulation with the
  conditions shown by the panel title, and the bottom panels (\textbf{A'} - \textbf{F'}) shows the $R_f$ concentration
 at four sites along the Dd cell boundary, with each neighboring pair separated by an angle of $\pi/2$.
\textbf{AA', BB', CC'} show a more directed Dd cell swimming when
guided by a single bacterium, whether with or without bacterial rheotaxis.
\textbf{DD', EE', FF'} show a more chaotic Dd cell swimming when surrounded by more 
bacteria.}
  \label{Fig.CellDirChange}
\end{figure}

We compare simulation results between systems with one bacterium only and systems with multi bacteria, the former group 
includes single bacterium with no external flow (figure \ref{Fig.CellDirChange}AA'), single bacterium with rheotaxis along the flow 
(figure \ref{Fig.CellDirChange}BB') and against the flow (figure \ref{Fig.CellDirChange}CC'),
and the latter multi bacteria group includes 6, 8 or 12 bacteria with no external flow (figure \ref{Fig.CellDirChange}DD', EE', FF'). In the initial stage of all the simulations, 
the chemoattractant signal has not diffused to the whole Dd cell, causing a random polarization of the cell. Therefore, PCI is high in a small 
period at the beginning stage. After the initial stage, the Dd cell is able to receive the diffusive chemoattractant signal. In the systems (no rheotaxis, 
rheotaxis along and against the flow) with only one bacterium, the averagely low PCI indicates that the Dd cell performs a more directed swimming 
toward the bacterium (figure \ref{Fig.CellDirChange}ABC). On the other hand, PCI is averagely high when more bacteria are present, indicating that the polarization 
decision of the Dd cell is affected by many body effect, and thus presents a chaotic behavior without the ability to effectively follow and catch a 
bacterium. Looking further into the dynamics of the chemotaxis guided amoeboid swimming (supplement movie and Figures \ref{Fig.mq7Ku6JH} - \ref{Fig.ssI3XJjA}), 
we find that the number multiplication of bacteria plays a similar role as the period multiplication as a road transition to the chaos
\cite{Cross1993Pattern}. Due to the Dd cell surrounded by multiple strong signal sources, 
the polarization of the Dd cell is frequently changed. 
In each of simulations in \ref{Fig.CellDirChange}, we track the $R_f$ concentrations at four sites along the Dd cell 
boundary, with each neighboring pair separated by an angle of $\pi/2$, and the results are given in the bottom panels in figure \ref{Fig.CellDirChange}\textbf{A' - F'}.
With only one bacterium in the system (figure \ref{Fig.CellDirChange}\textbf{A'B'C'}), $R_f  $ at the four sites are averagely low (mostly below $0.6$),
and it is clear that which site is at the front / rear, as its $R_f$ keeps the highest / lowest all the time (purple / orange line in figure \ref{Fig.CellDirChange}\textbf{A'B'C'}).
Such a clear difference in the $R_f $ level indicates a clear polarization of the Dd cell.
On the other hand, when more bacteria are presented in the system, $R_f$ levels are much higher (figure \ref{Fig.CellDirChange}\textbf{D'E'F'}),
and there is no clear high / low difference in the $R_f$ at the four sites due to the high level as well as noise, leading to the frequent change in the cell 
polarization as is shown by the PCI plots.
This causes the chaotic swimming pattern and reduces the cell's efficiency in catching bacteria. 
As pointed out above, this chaotic polarization behavior becomes more evident with more bacteria present in the system, which is verified by larger average PCIs in figure \ref{Fig.CellDirChange}F compared to figure \ref{Fig.CellDirChange}DE.

\section{Discussion}

While Dd amoeboid cell has long been well known as a model system for chemotaxis study on a crawling based motion, in recent year,
Dd cell swimming induced by different types of taxis including chemotaxis has become an emerging research area. A major modeling
challenge in this direction is the coupling of signaling dynamics and hydrodynamics. In this paper, we developed a minimal modeling 
framework to investigate the chemotaxis induced amoeboid cell swimming. Our model captures the interactions between
a Dd cell and bacteria, where both biochemical (chemotaxis signaling dynamics) and biomechanical (amoeboid swimming and bacterial rheotaxis) 
are considered. For the numerical computations, a complex analysis technique - \textit{mathematical amoeba model} 
\cite{shapere1987self,shapere1989geometry,avron2004optimal,Bouffanais2010hydrodynamics,wang2016computational}
is applied to solve the amoeboid dynamics in 2D viscous flows, associated with the finite volume method
based on the moving mesh Voronoi tessellation \cite{bottino2001computer,dillon2008single} to solve the RDC equation system.  
Our simulations results show that chemotaxis effectively guides Dd cell swimming, especially when less
bacteria are presented in the system, and bacterial rheotaxis may help a bacterium to escape from 
a predator Dd cell.

To better investigate the dynamics of chemotaxis induced amoeboid cell swimming, there are many aspects that our model 
can be developed. We discuss some major aspects in the following.

\textbf{Intracellular signaling dynamics induced amoeboid cell shape deformations.} In this paper, we used the 
2D \textit{mathematical amoeba model} (section \ref{Sec.CellDef-Swim}), which was greatly used in modeling
study of amoeboid cell swimming. However, the shape deformations in the mathematical amoeba model is prescribed.
In the future, we plan to develop an intracellular submodel for the amoeboid cell to capture how the membrane protein dynamics 
in response to extracellular stimuli generates excitable traveling waves of cell shape deformations. Several
modeling approaches to this directions have been developed, including models with a phenomenological description 
of membrane protein reaction-diffusion system that generates excitable dynamics of cell membrane deformation 
\cite{Campbell2017Computational,Campbell2020Computational} and a crawling based
chemotaxis induced amoeboid cell deformation and migration model \cite{Elliott2012Modelling}. We would like
to mention that the modeling framework developed in this paper is compatible with more complex 
cell deformations, using the computational method developed in \cite{wang2016computational}.

\textbf{Chemotaxis induced amoeboid swimming in confined space.} Currently in our model, we consider 
a swimming system in free space. However, amoeboid motion generally occurs close to surfaces, in small capillaries or in extracellular matrices of biological tissues. 
In addition, micro-organisms swim through permeable boundaries, cell walls, or microvasculature.
For example, flows are ubiquitous in human immune systems, blood vessels and microcirculation system, and are subjected to biological confinement by complex geometric structures. 
In particular, the effect of walls on motile micro-organisms has been a topic of increasingly active research. Recently, theoretical and modeling studies
have revealed complicated swimming trajectories with the confinement effects and simulation
predictions have been verified by experiments 
\cite{Wu2015Amoeboid,Wu2016Amoebid,Jana2012Paramecium,Ledesma2013Enhanced}.
In the future we will develop our model to study a swimming system in confined space.

\textbf{Hydrodynamic interactions and chemotaxis of bacteria.} In this paper we consider only a dilute suspension of bacteria,
and we neglect bacterium - bacterium and bacterium - Dd cell hydrodynamic interactions. Due to the large
size ratio of the Dd cell to a bacterium, the hydrodynamic effects generated by a bacterium should not affect
much of the Dd cell swimming dynamics, yet the hydrodynamic interactions between bacteria might play an important
role to bacterial swimming as well as the chemotaxis dynamics when the concentration of bacteria is higher. 
In recent years, both modeling and experimental studies reveal that in an active suspension
of bacteria, hydrodynamics affects bacterial collective motions with chemotaxis \cite{Sokolov2007Concentration,Lushi2018Nonlinear,Nejad2019Chemotaxis,Partridge2019Escherichia,Ryan2019Role}.
Another important future direction to our current modeling study would thus be to consider an active 
suspension of bacteria with hydrodynamic interactions.

In addition, it is well known that \textit{E. coli} also respond to chemotactic signals, either
produced by themselves or following local chemical gradients \cite{Lushi2012Collective, Ryan2019Role, Xue2011Travelling}.
Yet whether bacterial chemotaxis play a role in the Dd - \textit{E. coli} swimming system stays unclear. Would
bacterial chemotaxis help \textit{E. coli} to run away from the Dd cell is another interesting question to be considered and investigated,
on both experimental and modeling sides. In particular, two crucial questions should be addressed:
will the Dd send the signal to repel / attract the E. coli? With a large amount of \textit{E. coli} presented in the
system, how will the chemotaxis induced bacterial clustering alter the Dd - \textit{E. coli} interaction?

\ack The authors acknowledge partial support from the National Science Foundation Grant DMS-1951184 to QW.
Simulations were performed using the computer clusters and data storage resources of the HPCC at University of California, 
Riverside, which were funded by grants from NSF (MRI-1429826) and NIH (1S10OD016290-01A1).


 \appendix
 
 

\section{Goursat's formula and the conformal representation of the Dd cell}\label{Sec.2DLRNSwim}

Consider the Dd cell swimming in an incompressible Newtonian fluid of density $\rho$, viscosity $\mu$, and velocity $\mathbf{u}$,
the fluid dynamics is governed by the Navier-Stokes equations:
\begin{eqnarray}\label{Eq.NS1}
\rho \frac{\partial \mathbf{u}}{\partial t} + \rho (\mathbf{u} \cdot \nabla) \mathbf{u} = - \nabla p + \mu \Delta \mathbf{u} + \mathbf{f} \\ \label{Eq.NS2}
\nabla \cdot \mathbf{u} = 0
\end{eqnarray}
where the external force field $\mathbf{f}$ should vanish in a swimming problem as the cell
totally depends on self-propulsion. For the swimming Dd cell, let $L, U, \omega$ be the characteristic
scales of length, speed and frequency of shape deformations, respectively. We introduce two dimensionless
variables: Reynolds number $\textrm{Re} = \rho LU/\mu$ and the Strouhal number $\textrm{Sl} = \omega L/U$.
The Navier-Stokes equations (\ref{Eq.NS1}, \ref{Eq.NS2}) can be converted into dimensionless form:
\begin{eqnarray}\label{Eq.NS1-Scale}
\textrm{Re} \cdot \textrm{Sl} \frac{\partial \mathbf{u}}{\partial t} + \textrm{Re} (\mathbf{u} \cdot \nabla) \mathbf{u} = - \nabla p + \Delta \mathbf{u}  \\ \label{Eq.NS2-Scale}
\nabla \cdot \mathbf{u} = 0
\end{eqnarray}
The small size ($L$) and low speed ($U$) of cells leads to $\textrm{Re} \ll 1$, and in this low Reynolds number flow
regime cells move by exploiting the viscous resistance of the fluid. Dd celle have a typical length of
$L \sim 25 \mu$ m, swim at $U \sim 3 \  \mu$m/min and a period of shape deformation cycle $T \sim 1-2 $ min that 
gives the deformation frequency $\omega \sim 1$ min$^{-1}$ \cite{barry2010dictyostelium,van2011amoeboid}. Assume the medium is water ($\rho \sim 10^3 \textrm{kg}
 \cdot \textrm{m}^{-3}, \ \mu \sim 10^{-3} \textrm{Pa}\cdot \textrm{s}$), thus Re$\sim O (10^{-6})$ and Sl$\sim O (10^{-4})$.
For such a system, both inertial terms on the left hand side of equation (\ref{Eq.NS1-Scale}) can be neglected, and
the flow is governed by the Stokes equations:
\begin{eqnarray} \label{Eq.Stokes}
\Delta \mathbf{u} - \nabla p = \mathbf{0}, \quad \nabla \cdot \mathbf{u} = 0
\end{eqnarray}

In 2D the incompressibility condition $\nabla \cdot \mathbf{u} = 0$ in equation (\ref{Eq.Stokes}) can be satisfied 
by introducing a stream function $\Lambda (z, \overline{z};t)$, which is a real-valued scalar potential such that
$ u = \partial_y \Lambda - i \partial_x \Lambda$,
where $u \in \mathbb{C}$ is the complex representation of the velocity field $\mathbf{u}$, i.e., for $\mathbf{u} = (u_1, u_2)$,
$u = u_1 + i u_2$. Then the Stokes equations (\ref{Eq.Stokes}) imply that $\Lambda$ satisfies the biharmonic equation
$\Delta^2 \Lambda = 0$, whose general solution can be expressed by Goursat's formula \cite{muskhelishvili2013some}
\begin{eqnarray*}
\Lambda (z, \overline{z};t) = \mathfrak{R} [ \overline{z} \phi (z;t) + \chi (z;t)]
\end{eqnarray*}
where $\mathfrak{R} (\cdot)$ denotes the real part of the complex quantity; for any $t$, $\phi (z;t)$ and $\chi (z;t)$ are
analytic functions in the infinite fluid domain $\Omega(t)_{\textrm{Cell}}^C$, known as the \textit{Goursat functions}. Let 
$V (z, \overline{z};t)$ be the velocity boundary condition determined by the cell's shape deformations, 
with $z \in \partial \Omega_{\textrm{Cell}} (t)$, and we assume no-slip boundary condition along the cell's boundary,
then the 2D low Reynolds number swimming problem can be reduced to \cite{muskhelishvili2013some,shapere1989geometry}:
find analytic functions $\phi (z;t)$ and $\psi (z;t) = \chi' (z;t)$ in the fluid domain, such that
\begin{eqnarray}\label{Eq.2DLRN-BD-z}
\phi (z;t) - z \overline{\phi' (z;t)} - \overline{\psi (z;t)} = V (z, \overline{z};t) \qquad (z \in \partial \Omega_{\textrm{Cell}} (t))
\end{eqnarray}
Equation (\ref{Eq.2DLRN-BD-z}) will be referred to as the boundary condition constraint on the unknowns $\phi$
and $\psi$.
 
\begin{figure}[h]
\center
  \includegraphics[width=0.8\textwidth]{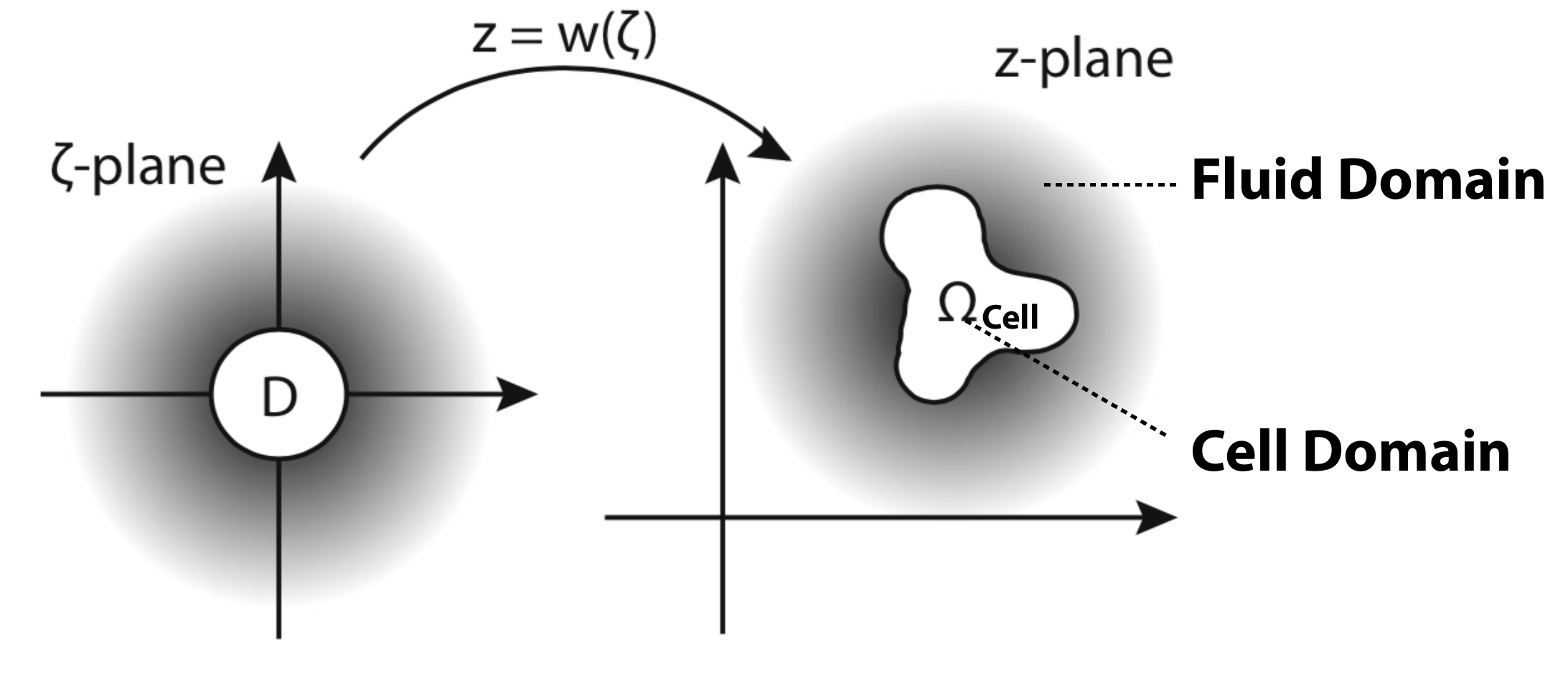}
  \caption{The conformal mapping $z=w(\zeta)$ from $\overline{\mathbb{C}} / \overline{D}$ to $\overline{\mathbb{C}} / \overline{\Omega}_{\textrm{Cell}}   $,
  with $\Omega_{\textrm{Cell}} $ and $\overline{\mathbb{C}} / \overline{\Omega}_{\textrm{Cell}} $ in the $z$-plane give the cell and fluid domains, respectively. Figure is
  reproduced from \cite{wang2016computational}.}
  \label{Fig.ConfMap}
\end{figure}

At time $t$, the Dd cell captures a simply connected bounded region $\Omega_{\textrm{Cell}}  (t)$ in the complex $z$-plane using
complex representation. Let $D = \{ \zeta \in \mathbb{C}: |\zeta| < 1 \}$ be the unit disk in the computational complex
$\zeta$-plane. The \textit{Riemann mapping theorem} ensures the existence of a single-valued analytic conformal
mapping $z = w (\zeta;t)$ which maps $\overline{\mathbb{C}} / \overline{D}$ one-to-one and onto 
$\overline{\mathbb{C}} / \overline{\Omega}_{\textrm{Cell}}  (t)$ -- the infinite fluid domain (Fig. \ref{Fig.ConfMap}), and preserves the correspondence
of infinity, i.e., $w (\infty;t) = \infty$. Therefore the shapes of the 
Dd cell when observed from the reference frame can be described by the $t$-family of conformal mapping $\{ z = w(\zeta;t) \}$.
In addition, $z = w(\zeta;t)$ always has its Laurent expansion of the form \cite{wang2016computational}:
\begin{eqnarray} \label{Eq.ConfMap}
z = w(\zeta;t) = \alpha_1 (t) \zeta + \alpha_0(t) + \frac{\alpha_{-1} (t)}{\zeta} + \frac{\alpha_{-2} (t)}{\zeta^2}
+ \cdots + \frac{\alpha_{-n} (t)}{ \zeta^n} + \cdots
\end{eqnarray}
where $\alpha_1 (t) \neq 0$ and $| \zeta |>1$. 
The $\alpha_0$ term gives the current location of the cell center, and the polarization can be indicated by the $\alpha_1$ term.
The Dd amoeboid cell boundary is thus given by
\begin{eqnarray*}
\partial \Omega_{\textrm{Cell}}  (t) = \{ z(t) = w(\sigma;t) | \sigma \in S^1 \}
\end{eqnarray*}
and the no-slip boundary condition can be written as
\begin{eqnarray*}
u \big( w (\sigma) \big) (t) = \frac{\partial}{\partial t} w (\sigma ; t)
\end{eqnarray*}

Let $\phi (z,t)$ and $\psi (z,t)$ be the solutions to the boundary condition constraint given by equation (\ref{Eq.2DLRN-BD-z}),
and let 
\begin{eqnarray*}
\Phi(\zeta;t) = \phi (w (\zeta;t);t), \ \Psi (\zeta;t) = \psi (w (\zeta;t);t) \quad (| \zeta| \geq 1)
\end{eqnarray*}
thus $\Phi(\zeta;t)$ and $\Psi (\zeta;t)$ are functions on the $\zeta$-plane which are analytic on $\mathbb{C} / \overline{D}$ and continuous on
$\mathbb{C} / D$ at any time $t$. With the conformal mapping representation, equation (\ref{Eq.2DLRN-BD-z}) can be pulled back to the 
$\zeta$-plane as a function on the unit disk $S^1$:
\begin{eqnarray}\label{Eq.2DLRN_BD-zeta}
\Phi (\sigma) - \frac{w (\sigma)}{\overline{w'(\sigma)}} \overline{\Phi' (\sigma)} - \overline{\Psi (\sigma)} = V (\sigma) \qquad (\sigma \in S^1)
\end{eqnarray}
 
In general, the $\zeta^{-n}$ term with $n>0$ in the conformal mapping equation (\ref{Eq.ConfMap}) gives $n+1$ angles
along a round periphery of the swimmer. To approximate a real swimming Dd cell shape, we can first approximate the cell
boundary by a $N$-polygon, then obtain the corresponding conformal mapping using the Schwarz-Christoffel formula then 
truncate its Laurent expansion leaving a finite number of terms, from which we can solve equation (\ref{Eq.2DLRN_BD-zeta})
using the computational Muskhelishvili's method, as prescribed in \cite{wang2016computational}.
 
Here we adopt a simple form of this mathematical amoeba model, where the shape of cell is in the form of
\begin{eqnarray}\label{Eq.AmoebaConfMap-2Term}
w(\zeta;t) = e^{i \theta (t)} \Big[ r(t)  \zeta  + \frac{\eta_{-1} (t)}{\zeta} + \frac{\eta_{-2} (t)}{\zeta^2} \Big] + Z_{\textrm{Dd}}(t)
\end{eqnarray}
where 
\begin{itemize}
\item $r(t) \in \mathbb{R}$  controls the cell size. When the cell is at rest and in a circular shape ($\eta_{-1}  = \eta_{-2} =0$), $r$ 
gives the cell radius, which we denote by $r_0$. We assume that as the cell swims, there is no material exchange between the cell
and the surrounding fluid, which validates the no-slip boundary condition. In 2D it naturally becomes the area conservation constraint,
described by the following equation:
\begin{eqnarray*}
\textrm{Area} (t) = \frac{1}{2} \mathfrak{I} \oint \overline{w} dw \equiv \pi r_0^2
\end{eqnarray*}
where $\mathfrak{I}$ denotes the imaginary part of the complex quantity. With $w$ given by equation (\ref{Eq.AmoebaConfMap-2Term}),
the area conservation constraint can be reduced to:
\begin{eqnarray}\nonumber
\textrm{Area} (t) &=& \pi  \Big( r^2 (t)- | \eta_{-1} (t) |^2 - 2 | \eta_{-2} (t) |^2  \Big) \equiv \pi r_0^2 \\ \label{Eq.AreaConsv}
\Rightarrow & & r (t) =  \sqrt{r_0^2 + | \eta_{-1} (t) |^2 + 2 | \eta_{-2} (t) |^2 }
\end{eqnarray}
\item $ \theta (t) \in \mathbb{R}$ gives the cell polarization that will be determined by signaling sensing dynamics as 
discussed in the Section~\ref{Sec.Pol-Def}. In particular, we assume that the polarization $\theta$ is determined at the beginning of
a swimming cycle and will not change during the cycle, thus $\theta (nT+t) \equiv \theta (nT)$ during the $n$th swimming cycle with $t \in [0,T)$.
\item $\eta_{-1} (t), \eta_{-2} (t) \in \mathbb{R}$ give the shape deformations. They are real-valued 
functions so that the cell swims in a straight line once the polarization $\theta$ is determined from signaling dynamics.
\item $Z_{\textrm{Dd}}(t)$ gives the location of the cell, while $U_{\textrm{Dd}}(t) = \dot{Z}_{\textrm{Dd}}(t)$ 
gives the velocity of the cell which is computed through cell-fluid interaction discussed as follows.
\end{itemize}

In the case that the conformal mapping equation (\ref{Eq.ConfMap}) includes only the $\zeta^{-1}, \zeta^{-2}$ in the
negative order terms in its Laurent expansion, the solution to equation (\ref{Eq.2DLRN_BD-zeta}) is given by \cite{shapere1989geometry,avron2004optimal}:
\begin{eqnarray*}
\Phi(\zeta) &=& \dot{Z}_{\textrm{Dd}} + \frac{\dot{\alpha}_{-1}}{\zeta} + \frac{\dot{\alpha}_{-2}}{\zeta^2}  \\
\Psi (\zeta) &=& - \frac{\overline{\dot{\alpha}_1}}{\zeta} + 
\frac{\overline{\alpha_{-2}} + \frac{\overline{\alpha_{-1}}}{\zeta} + \frac{\overline{\alpha_1}}{\zeta^3}}{\alpha_1 - \frac{\alpha_{-1}}{\zeta^2} - \frac{2 \alpha_{-2}}{\zeta^3}}
\Big( \dot{\alpha}_{-1} + \frac{2 \dot{\alpha}_{-2}}{\zeta} \Big)
\end{eqnarray*}
In particular, with $w$ given by equation (\ref{Eq.AmoebaConfMap-2Term}) and that $\theta (nT+t) \equiv \theta (nT)$ 
during each swimming stroke, we have
\begin{eqnarray}\label{Eq.2DLRN-Phi}
\Phi(\zeta) &=& e^{i \theta} \Big(  \frac{\eta_{-2} \dot{\eta}_{-1}}{r} + \frac{\dot{\eta}_{-1}}{\zeta} + \frac{\dot{\eta}_{-2}}{\zeta^2}  \Big) \\ \label{Eq.2DLRN-Psi}
\Psi (\zeta) &=& e^{-i \theta} \Big[ - \frac{\dot{r}}{\zeta}  + \frac{\eta_{-2} + \frac{\eta_{-1}}{\zeta} + \frac{r}{\zeta^3}}{r - \frac{\eta_{-1}}{\zeta^2} - \frac{2 \eta_{-2}}{\zeta^3}} 
\Big( \dot{\eta}_{-1} + \frac{2 \dot{\eta}_{-2}}{\zeta} \Big) \Big] 
\end{eqnarray}
where
\begin{eqnarray} \nonumber
U_{\textrm{Dd}} &=& \dot{Z}_{\textrm{Dd}} = e^{i \theta}  \frac{\eta_{-2} \dot{\eta}_{-1}}{r} \\ \label{Eq.De-r}
\dot{r} &=& \frac{\eta_{-1} \dot{\eta}_{-1} + 2 \eta_{-2} \dot{\eta}_{-2}}{\sqrt{r_0^2 + \eta_{-1}^2 + 2 \eta_{-2}^2}}
\end{eqnarray}
The velocity field of the surrounding flow is thus given by the formula
\begin{eqnarray}\label{Eq.FluidVelocity}
u (\zeta, \overline{\zeta}) = \Phi (\zeta) - \frac{w (\zeta)}{\overline{w'(\zeta)}} \overline{\Phi' (\zeta)} - \overline{\Psi (\zeta)} \qquad (|\zeta| \geq 1)
\end{eqnarray}
$u (\zeta, \overline{\zeta})$ induces the velocity function $u_z$ on the $z$-plane:
\begin{eqnarray*}
u_z (z) = u \big( \zeta= w^{-1} (z) \big)
\end{eqnarray*}
With $z=w(\zeta)$ given by equation (\ref{Eq.AmoebaConfMap-2Term}), by Lagrange inversion formula, we have
\begin{eqnarray}\label{Eq.w-Inverse}
\frac{1}{\zeta} = \frac{e^{i \theta} r}{z - Z_{\textrm{Dd}}} +  \frac{e^{3 i \theta} r^2 \eta_{-1}}{ (z - Z_{\textrm{Dd}})^3} 
+  \frac{e^{4 i \theta} r^3 \eta_{-2}}{ (z - Z_{\textrm{Dd}})^4} + O \Big(  \frac{1}{ (z - Z_{\textrm{Dd}})^5} \Big)
\end{eqnarray}
We use equation (\ref{Eq.w-Inverse}) to obtain an approximate for $u_z (z) $:
\begin{eqnarray}\label{Eq.uz}
u_z (z) = u \big( \zeta= w^{-1} (z) \big) \sim u \Big(  \big(  \frac{e^{i \theta} r}{z - Z_{\textrm{Dd}}} +  \frac{e^{3 i \theta} r^2 \eta_{-1}}{ (z - Z_{\textrm{Dd}})^3} 
+  \frac{e^{4 i \theta} r^3 \eta_{-2}}{ (z - Z_{\textrm{Dd}})^4} \big)^{-1} \Big) \qquad \qquad 
\end{eqnarray}
In the following discussion for simplicity we use $u (z)$ to denote $u_z (z)$.


\section{Signaling induced Dd cell polarization and shape deformations} \label{Sec.Pol-Def}
 
We assume that the Dd cell undergoes shape deformations in response to signal gradient, with
each swimming stroke lasting for a period of $[0, T]$ and consisting of three phases: (I) \textit{polarization}, (II) \textit{swimming},
(III) \textit{relaxation}.

\begin{enumerate}
\item \textit{Polarization.} At the beginning of each cycle, the amoeba is of a circular 
shape with no polarization, i.e., $w(\zeta) = r_0 \zeta + Z_{\textrm{Dd}} $.
The $R_f$ concentration is numerically evaluated at each $ \mathbf{w}'_j$.
If 
\begin{eqnarray*}
 \max_{1 \leq j \leq N_r} \{ R_f (\mathbf{w}'_j) \}  -  \min_{1 \leq j \leq N_r} \{ R_f (\mathbf{w}'_j) \}  < \varepsilon_R
\end{eqnarray*}
for some threshold $\varepsilon_R > 0$, we randomly choose a polarization direction $\theta_T  $. Otherwise,
we find the node $ \mathbf{w}'_K$ that has 
the maximum $R_f$ value: $R_f (\mathbf{w}'_K) = \max_{1 \leq j \leq N_r} \{ R_f (\mathbf{w}'_j) \} $. Then 
$\theta_K$ corresponding to $\mathbf{w}'_K$ defines the polarization of the following stroke:
\begin{eqnarray*}
\theta_T = \theta (t) \big|_{t \in [0,T]} \equiv \theta_K = - i \ln \Big( w^{-1} ( \mathbf{w}'_K) \Big)
\end{eqnarray*}
Once the polarization $\theta_T$ is determined, in a short period $[0,T_P] \subset [0,T]$, the Dd cell stretches itself from 
a circular shape to an ellipse with its semi-major axis lying along the $\theta_T$ direction: 
\begin{eqnarray}\label{Eq.CellShape-Pol}
w (\zeta;t) = e^{i \theta_T} \big[ r(t) \zeta + \frac{\eta_{-1} (t)}{\zeta} \big] + Z_{\textrm{Dd}} (t), \qquad t \in [0, T_P]
\end{eqnarray}
where we can choose $\eta_{-1} (t) $ to be the following linear function:
\begin{eqnarray*}
\eta_{-1} (t) = \frac{t}{T_P} \overline{\eta}, \qquad t \in [0, T_P]
\end{eqnarray*}
such that $\eta_{-1} (0) = 0$ gives the circular shape, and $\eta_{-1} (T_P) = \overline{\eta}$ gives an ellipse
with semi-major and semi-minor axis lengths $r(T_P) +\overline{\eta}$ and $r(T_P) - \overline{\eta}$, respectively.
The deformation in this stage will not result in net translation according to the \textit{scallop theorem} \cite{purcell1977life},
thus $Z_{\textrm{Dd}} (t) \equiv Z_{\textrm{Dd}} (0)$ for $t \in [0, T_P]$.

\item \textit{Swimming.} After polarized, the Dd cell undergoes the following
shape deformations in a time interval $[T_P, T_P + T_S] \subset [0,T]$ which leads to active swimming along the polarization direction
as is discussed in Section~\ref{Sec.2DLRNSwim}:
\begin{eqnarray}\label{Eq.CellShape-Swim}
w (\zeta; t) = e^{i \theta_T} \big[ r(t) \zeta + \frac{\overline{\eta} \cos (2 \pi \tau)}{\zeta} -  \frac{\overline{\eta} \sin (2 \pi \tau)}{\zeta^2} \big] + Z_{\textrm{Dd}} (t)
\end{eqnarray}
where $\tau = (t-T_P)/T_S$.

\item  \textit{Relaxation.} After the amoeboid cell returns to the ellipse shape at $t = T_P + T_S$, in the followed $[ T_P + T_S, 2T_P+T_S ]\subset [0,T]$ period
the Dd cell undergoes the reversed shape deformations as defined in equation (\ref{Eq.CellShape-Pol}) but with
\begin{eqnarray*}
\eta_{-1} (t) = \frac{T - t}{T_P} \overline{\eta}, \qquad t \in [ T_P + T_S, T]
\end{eqnarray*}
where $T_P$, $T_S$ should satisfy $2T_P+T_S = T$.
The Dd cell returns to the initial circular shape by the end of the stroke $t = T$.
Again, the deformation in this stage do not result in net translation, i.e., 
$Z_{\textrm{Dd}} (t) \equiv Z_{\textrm{Dd}} (T_P + T_S)$ for $t \in [T_P + T_S, T]$.
A full swimming stroke is thus completed.

\end{enumerate}


\section{Nondimensionalization of the system}\label{Sec.Non-Dim-Supp}

We nondimensionalize the system, with $T$ (the duration of a Dd cell swimming cycle) and $r_0$ (the Dd cell radius when the cell is at rest) 
the characteristic spatial and temporal scales, and $R_{\max}$ the characteristic length concentration scale:
\begin{eqnarray*}
t^* = \frac{t}{T}, \ \mathbf{x}^* = \frac{\mathbf{x}}{r_0}, \ z^* = \frac{z}{r_0}, \ f^* = \frac{f}{R_{\max}  /  r_0} ,  \\ 
R_f^{0,*} = \frac{R_f^0}{R_{\max}},  \ R_f^* = \frac{R_f}{R_{\max}}, \ 
d W^* = \frac{dW}{\sqrt{T}}  
\end{eqnarray*}
When substitute into the model equations, bacteria motion equation (Eq (1) in the main text) becomes:
\begin{eqnarray}\label{Eq.Scale-BacMotion}
 d \mathbf{x}^*_n  = \mathbf{u}^*  d t^* + d \mathbf{X}_n^*
\end{eqnarray}
The RDC equations (Eqs (3-5) in the main text) become:
\begin{eqnarray}\label{Eq.Scale-fol}
\frac{\partial f^* }{\partial t^*} &=& D^* \Delta^* f^*- \mathbf{u}^*   \cdot \nabla^* f^* 
 +  a^* \int \sum_{n=1}^{N_B}   \delta (\mathbf{x}^* - \mathbf{x}_n^*) d \mathbf{x}^*  \\ \label{Eq.Scale-fol_bd_cell}
D^* \mathbf{n} \cdot   \nabla^* f^* &=& - k_+^* f^* R_f^{0,*} + k_-^* R_f^* \\  \label{Eq.Scale-fAR}
\frac{\partial R_f^* }{\partial t^*} &=& k_+^* f^* R_f^{0,*}  - k_-^* R_f^*
- \gamma^* R_f^* + \varsigma^* R_f^* \frac{dW^*}{dt^*}
\end{eqnarray}
where
\begin{eqnarray*} 
  \Delta^* = r_0^2 \Delta, \ \nabla^* = r_0 \nabla, \ D^* = \frac{DT}{r_0^2}, \ k_+^* = \frac{k_+ R_{\max}  T}{r_0},  \ k_-^* =  k_- T, \\
   a^* = \frac{ar_0  T}{R_{\max}},  \ \gamma^* = \gamma T, \
   \varsigma^*  = \varsigma  \sqrt{T} 
\end{eqnarray*}
The cell shape equation (\ref{Eq.AmoebaConfMap-2Term}) and the area conservation constraint equation (\ref{Eq.AreaConsv}) become
\begin{eqnarray}\label{Eq.Scale-AmoebaConfMap-2Term}
w^*  &=& e^{i \theta  } \Big[ r^*   \zeta  + \frac{\eta_{-1}^*  }{\zeta} + \frac{\eta_{-2}^*  }{\zeta^2} \Big] + Z^*_{\textrm{Dd}}  \\   \label{Eq.Scale-AreaConsv}
r^*  &=&  \sqrt{1 + | \eta_{-1}^*   |^2 + 2 | \eta_{-2}^*  |^2 }
\end{eqnarray}
with 
\begin{eqnarray*}
& & r^* (t^* ) = \frac{r(t^* )}{r_0}, \ \eta_{-1}^* (t^* ) = \frac{\eta_{-1}(t^* )}{r_0}, \  \eta_{-2}^* (t^* ) = \frac{\eta_{-2}(t^* )}{r_0} \\
& & U_{\textrm{Dd}}^* = \frac{d Z^*_{\textrm{Dd}}}{d t^*} = e^{i\theta}  \frac{\eta^*_{-2} }{r^*} \frac{d \eta_{-1}^*}{d t^*}
\end{eqnarray*}
The fluid velocity field equation (\ref{Eq.FluidVelocity}) becomes:
\begin{eqnarray}\label{Eq.Scale-FluidVelocity}
u^* (\zeta, \overline{\zeta}) = \Phi^* (\zeta) - \frac{w^* (\zeta)}{\overline{{w^*}'(\zeta)}} \overline{{\Phi^*}' (\zeta)} - \overline{\Psi^* (\zeta)} \qquad (|\zeta| \geq 1)
\end{eqnarray}
where the non-dimensional Goursat's functions are:
\begin{eqnarray*}
\Phi^* = \frac{T}{r_0} \Phi, \quad \Psi^* = \frac{T}{r_0} \Psi
\end{eqnarray*}
The inverse of the conformal mapping equation (\ref{Eq.w-Inverse}) becomes
\begin{eqnarray}\label{Eq.Scale-w-Inverse}
\frac{1}{\zeta} = \frac{e^{i \theta} r^*}{z^* - Z_{\textrm{Dd}}^*} +  \frac{e^{3 i \theta} r^{*2} \eta_{-1}^*}{ (z^* - Z_{\textrm{Dd}}^*)^3} 
+  \frac{e^{4 i \theta} r^{*3} \eta_{-2}^*}{ (z^* - Z_{\textrm{Dd}}^*)^4} + O \Big(  \frac{1}{ (z^* - Z_{\textrm{Dd}}^*)^5} \Big)
\end{eqnarray}
From equations (\ref{Eq.2DLRN-Phi}, \ref{Eq.2DLRN-Psi}) we have
\begin{eqnarray}\label{Eq.Scale-2DLRN-Phi}
\Phi^*(\zeta) &=& e^{i \theta} \Big(  \frac{\eta^*_{-2} }{r^*} \frac{d \eta_{-1}^*}{d t^*}+ \frac{1}{\zeta} \frac{d \eta_{-1}^*}{d t^*} + \frac{1}{\zeta^2} \frac{d \eta_{-2}^*}{d t^*} \Big) \\ \label{Eq.Scale-2DLRN-Psi}
\Psi^* (\zeta) &=& e^{-i \theta} \Big[ - \frac{1}{\zeta}  \frac{d r^*}{d t^*} + \frac{\eta^*_{-2} + \frac{\eta^*_{-1}}{\zeta} + \frac{r^*}{\zeta^3}}{r^*- \frac{\eta^*_{-1}}{\zeta^2} - \frac{2 \eta^*_{-2}}{\zeta^3}} 
\Big(\frac{d \eta^*_{-1}}{d t^*} + \frac{2  }{\zeta} \frac{d \eta^*_{-2}}{d t^*} \Big) \Big] 
\end{eqnarray}
where from equation (\ref{Eq.De-r})
\begin{eqnarray*}
\frac{d  r^*}{d t^*} &=& \frac{1}{ \sqrt{1 + | \eta_{-1}^*   |^2 + 2 | \eta_{-2}^*  |^2 }} \Big( \eta_{-1}^* \frac{\eta_{-1}^*}{d t^*} +  \eta_{-2}^* \frac{\eta_{-2}^*}{d t^*}  \Big)
\end{eqnarray*} 
Finally we also give the expressions of ${w^*}'$ and ${\Phi^*}'$ that will be used in calculating the fluid velocity field equation (\ref{Eq.Scale-FluidVelocity}):
\begin{eqnarray*}
\frac{d w^*}{d \zeta} &=& e^{i \theta} \Big( r^* - \frac{\eta_{-1}^*}{\zeta^2} - \frac{2 \eta_{-2}^*}{\zeta^3} \Big) \\
\frac{d \Phi^*}{d \zeta} &=&- e^{i \theta} \Big( \frac{1}{\zeta^2}  \frac{d \eta_{-1}^*}{d t^*} + \frac{2}{\zeta^3}  \frac{d \eta_{-2}^*}{d t^*} \Big)
\end{eqnarray*}


\section{Local re-mesh of the fluid nodes near the Dd cell}\label{Sec.Remesh-Supp}

To avoid clustering of fluid nodes near the cell boundary due to the Dd cell's large deformations, we apply two
types of local re-mesh of the fluid nodes $\{ \mathbf{w}_i \}$. First, at each time step, the closest layer of 
fluid nodes to the Dd cell is generated from the current conformal mapping $w$ (equation (\ref{Eq.Scale-AmoebaConfMap-2Term}),
we omit the $*$ sign for dimensional quantities for simplicity and also in the following discussions)
by taking equally spaced nodes along the contour generated by 
$$ w = e^{i \theta} \big[ (1 + \epsilon) \sigma + \frac{\eta_{-1}}{\sigma} + \frac{\eta_{-2}}{\sigma^2} + Z_{Dd} \big], \qquad \sigma \in S^1$$
for a small number $\epsilon \approx 0.3$. In addition, at the beginning of every cycle ($t = nT$), we re-generate fluid nodes in a 
local region near the cell boundary $ \{ \mathbf{x} | \| \mathbf{x} - \mathbf{x}_{\textrm{Dd}} \| < 3  \} $ in a cubic lattice --
the same as we initially generate the fluid mesh at $t=0$, where
$ \mathbf{x}_{\textrm{Dd}}$ is the Dd cell center position, and the distance is after nondimensionalization, i.e.,
the Dd cell has radius$=1$ at rest. After each re-mesh, the $f$ values at each newly generated node
is taken to be the average of the three closest neighboring nodes from the old mesh. Finally we remove the old
nodes in the re-meshed region.

 
 \section{Parameters}
 
 The parameter values used in our simulations are listed in the table below.
Default values are for dimenionless quantities.
    \begin{center}
\captionsetup{width=\textwidth}
     \begin{longtable}{p{2.5cm}p{2.5cm}p{4cm}p{3cm}}
    \label{Tab.Para}
    \endfirsthead
    \endhead
    \hline
    Parameters &  Unit & Non-dimensionalization & Default Value  \\  
                   \hline
     \multicolumn{4}{c}{Bacterial motions} \\
     \hline
$\delta_J$    & m  &  $\delta_J^* = \delta_J / r_0$  &  $0.05$ \\ 
$M$ & m/s &  $M^* =TM/r_0 $  &  $1$ \\ 
 \hline
         \multicolumn{4}{c}{Chemotaxis signaling dynamics}  \\
     \hline
     $D$ & m$^2/$s &  $D^* =DT/r_0^2 $  &  $1$ \\ 
          $a$ & mol/m$^2$s &  $a^* =a r_0  T/ R_{\max} $  &  $1$ \\ 
   $k_+$ & m$^2$/mol$\cdot$s  &  $ k_+^* = k_+ R_{\max} T / r_0 $  &  $1$ \\ 
     $k_-$ & 1/s  &  $ k_-^* = k_- T $  &  $0$ \\ 
$\gamma$ & 1/s  &  $ \gamma^* = \gamma T $  &  $0.2$ \\ 
$\varsigma $ & 1/$\sqrt{\textrm{s}}$  &  $ \varsigma^*  = \varsigma  \sqrt{T} $  &  $0.1$ \\ 
       \hline
 \caption{Parameter values.} 
     \end{longtable}
 
\end{center}

The characteristic scales we used to non-dimensionalize the system are:
\begin{itemize}
\item $T$ - the duration of a Dd cell swimming cycle;
\item $r_0$ - the Dd cell radius when the cell is at rest;
\item $R_{\max}$ - characteristic length concentration of $R_f$.
\end{itemize} 
From \cite{barry2010dictyostelium, van2011amoeboid}, we have an estimate of $T$ about $1 - 1.5$ minute;
while for the cell size, only the length and width of a slender cell are available: $\sim 25 \mu m$ long and $\sim 6 \mu m$
wide as reported in \cite{van2011amoeboid}, and $\sim 22 \mu m$ long and $\sim 4 \mu m$
wide as reported in \cite{barry2010dictyostelium}, from which we can estimate that $r_0$ should be at the scale of 
$O (10) \  \mu m$. We are not aware of available data on $R_{\max}$.

If $T, \ r_0$ and $R_{\max}$ can be measured in lab, estimates of all quantities in the system can be 
obtained from Table \ref{Tab.Para}. For example, in our simulations we take $D^* \in [0.2, 1], \ \delta_J^* \in [0.02, 0.1]$,
with an estimate of $T = 1 \textrm{ min} = 60 \textrm{ s}$, $r_0  \sim 10 \mu \textrm{m} = 10^{-5} \textrm{ m}$, then we 
estimate the range of $D$ and $\delta_J$ should be about:
$$ D \sim O (10^{-13} - 10^{-12}) \  \textrm{m}^2/\textrm{s}, \quad  \delta_J \sim O (10^{-7} - 10^{-6}) \ \textrm{m}$$

 
 \section{Supplement figures}
 
 \begin{figure}[h]
\center
  \includegraphics[width=0.95\textwidth]{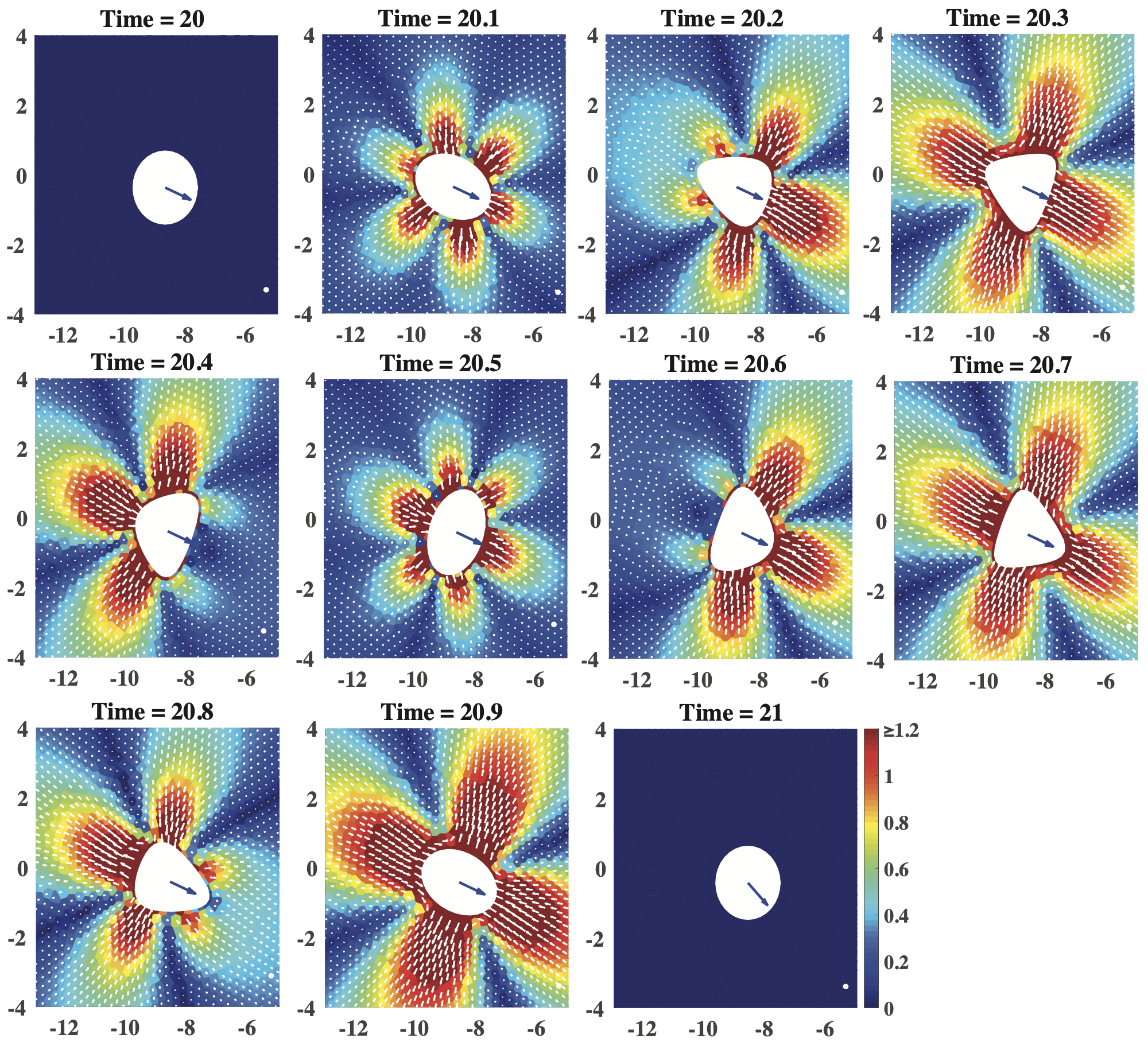}
  \caption{Time lapse snapshots of the fluid velocity profile during one Dd cell swimming cycle, where the heatmap shows 
  the amplitude of the fluid velocity, and the white arrows show the fluid velocity directions. Notice that the flow directions keep changing
 with the cell deformation. For example, the flow is toward the cell from the back at Time=20.4, while at Time=20.7, the flow is away from the cell from the back.}
  \label{Fig.BaUNqZ68-Cycle}
\end{figure}

 \clearpage
\newpage
 \begin{figure}[h]
\center
  \includegraphics[width=0.95\textwidth]{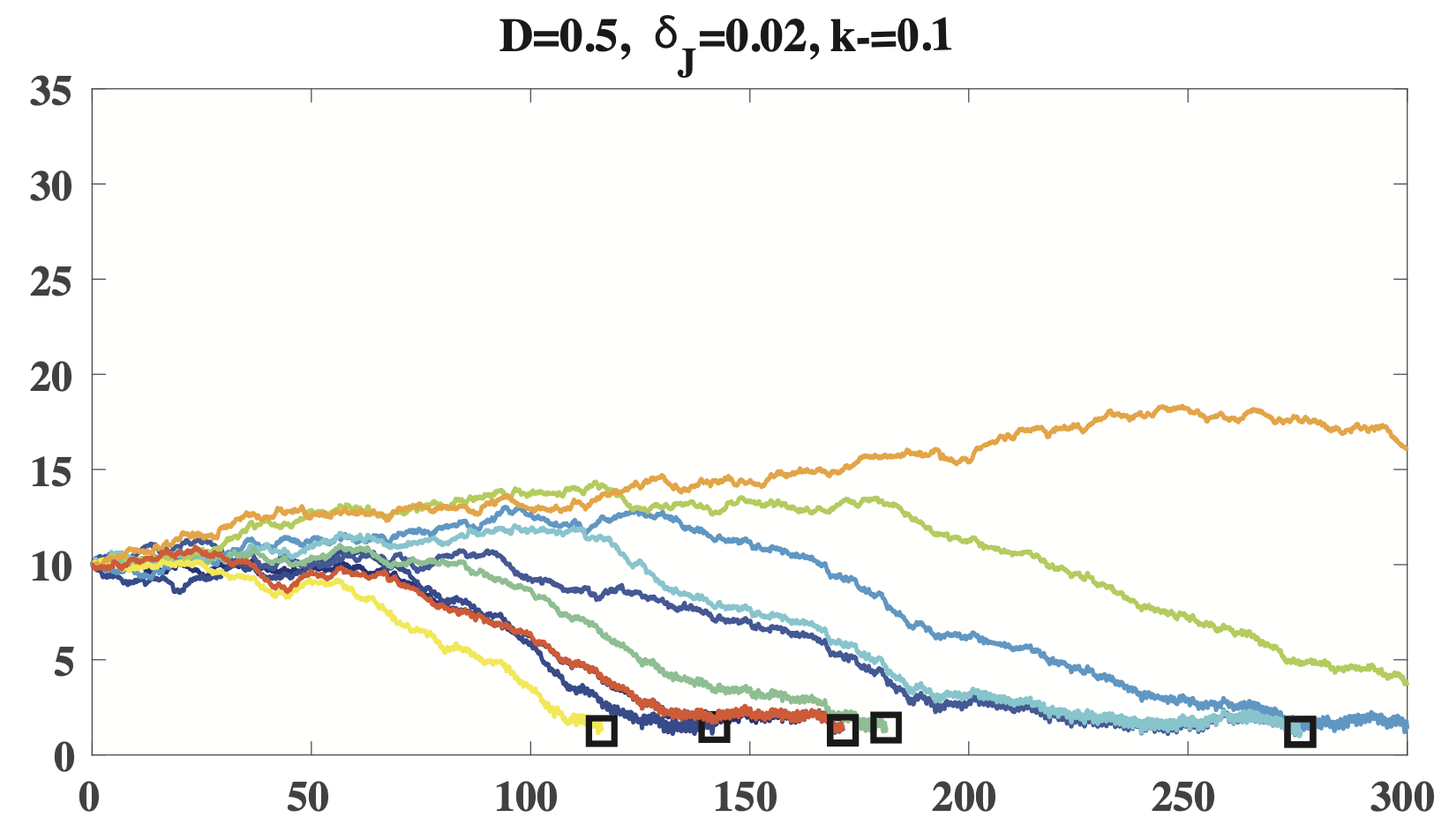}
  \caption{Distance between the Dd cell center and the bacterium with $D=0.5, \ \delta_J = 0.02$ and $k_- = 0.1$. No
  bacterial rheotaxis.}
  \label{Fig.CellDist_Km}
\end{figure}

 \clearpage
\newpage
 \begin{figure}[h]
\center
  \includegraphics[width=0.95\textwidth]{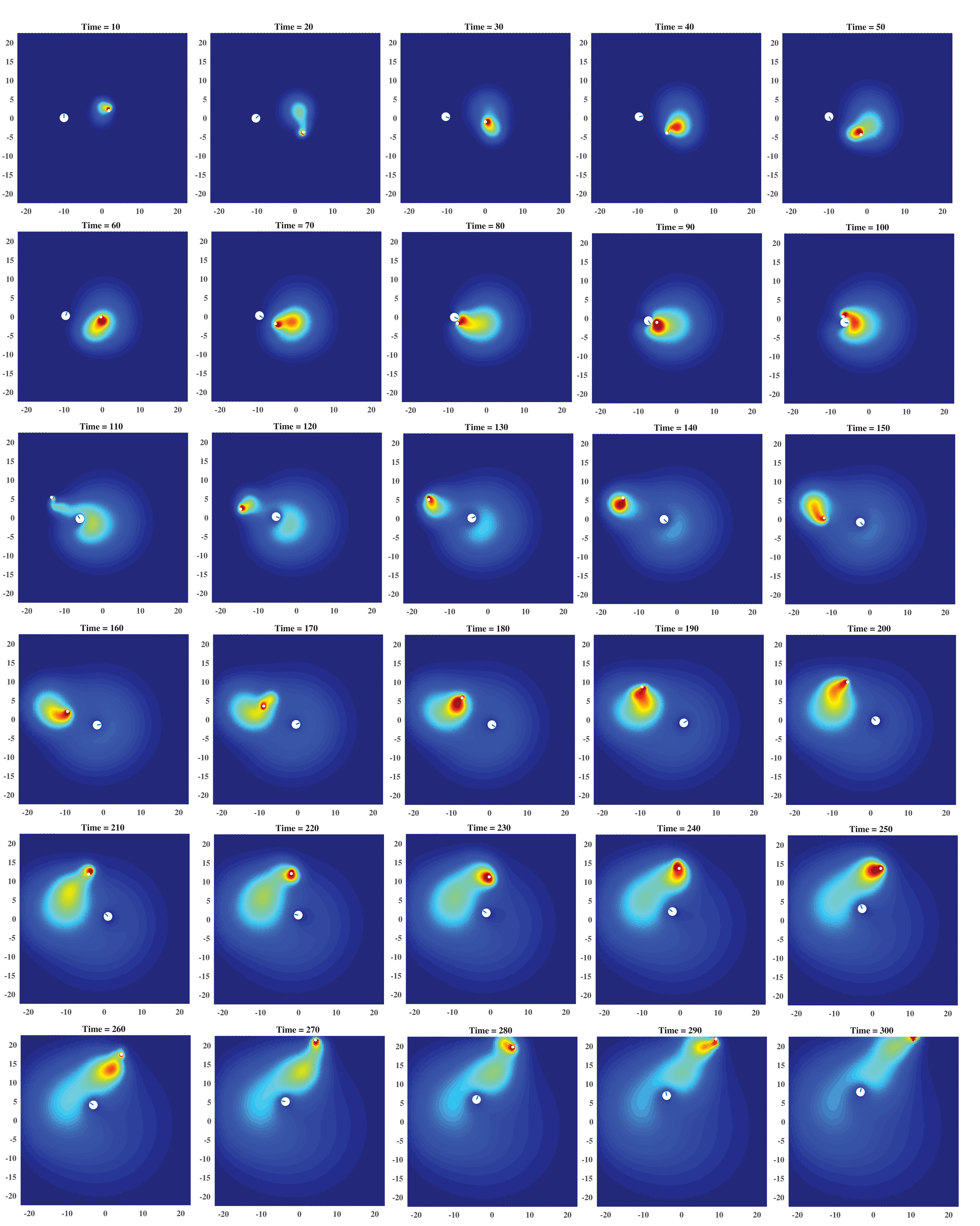}
  \caption{Time lapse snapshots of a typical simulation with $D = 0.2, \ \delta_J = 0.1$ and no bacterial rheotaxis. The color scale
  is the same as main figure 3C.}
  \label{Fig.Mcf7QdMF}
\end{figure}

\clearpage
\newpage
 \begin{figure}[h]
\center
  \includegraphics[width=0.95\textwidth]{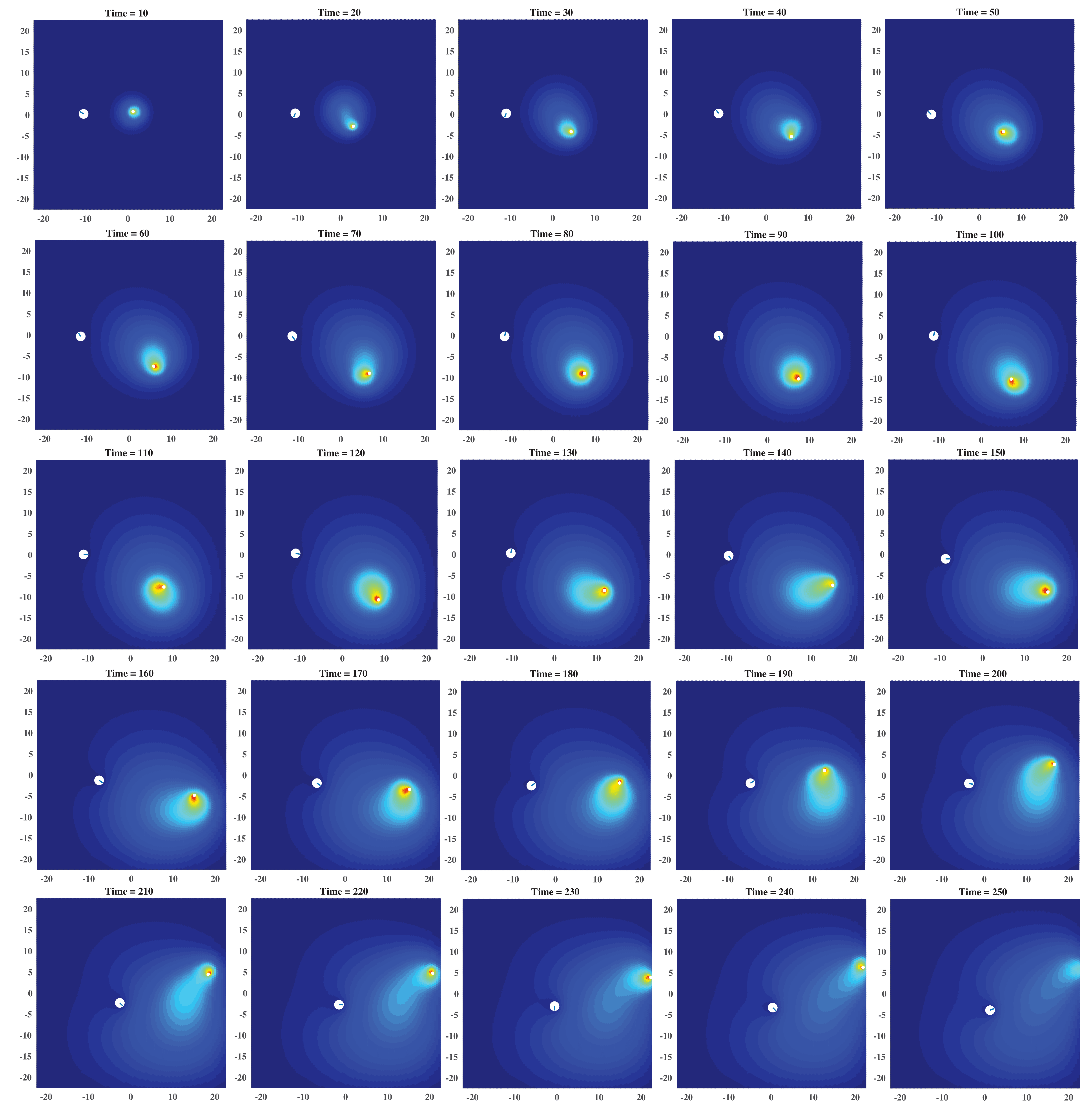}
  \caption{Time lapse snapshots of a typical simulation with $D = 0.5, \ \delta_J = 0.1$ and no bacterial rheotaxis. The color scale
  is the same as main figure 3C.}  
  \label{Fig.Dpt5j1e1}
\end{figure}

\clearpage
\newpage
 \begin{figure}[h]
\center
  \includegraphics[width=0.95\textwidth]{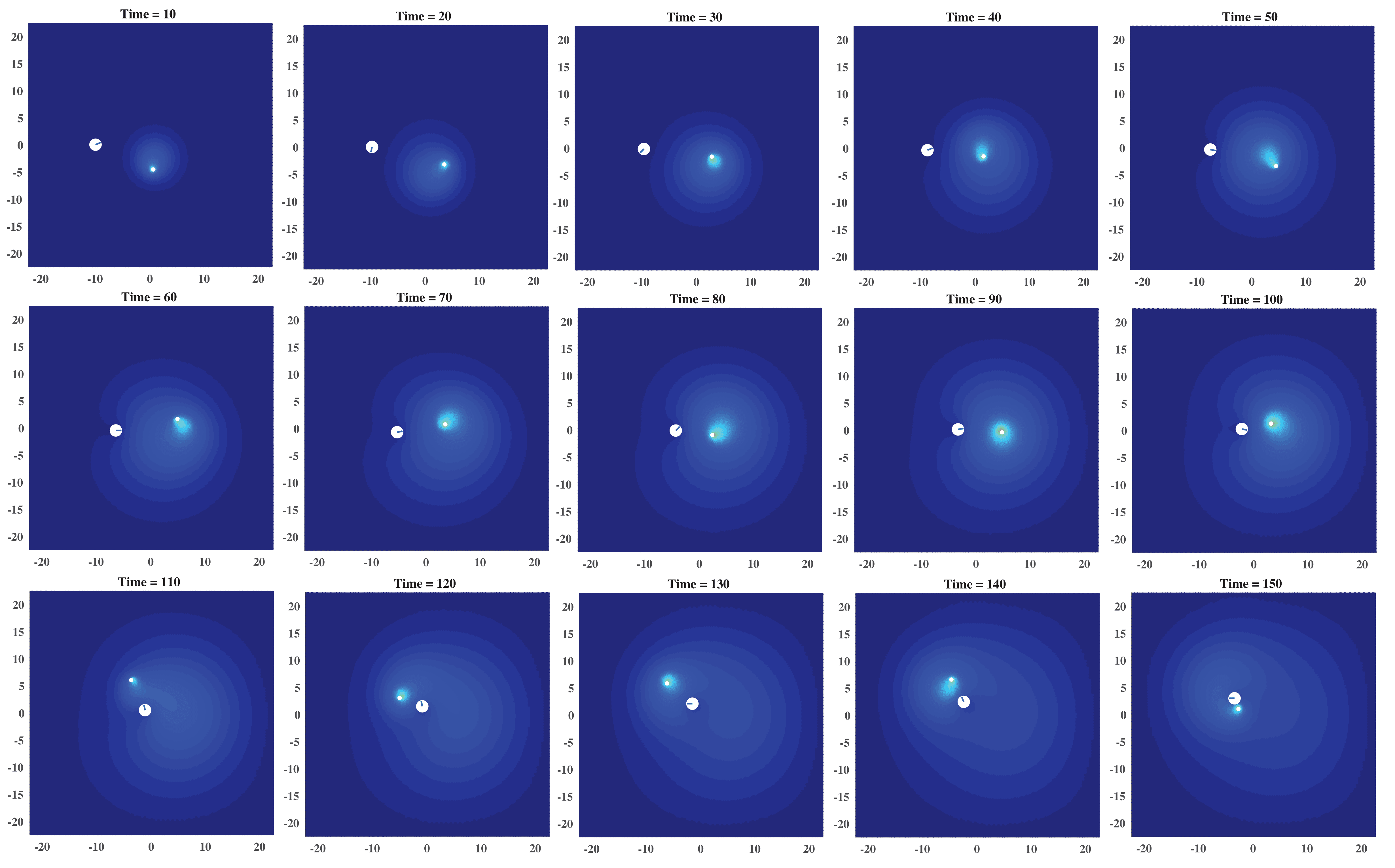}
  \caption{Time lapse snapshots of a typical simulation with $D = 1, \ \delta_J = 0.1$ and no bacterial rheotaxis. The color scale
  is the same as main figure 3C.} 
  \label{Fig.HuyRtX3j}
\end{figure}

\clearpage
\newpage
 \begin{figure}[h]
\center
  \includegraphics[width=0.95\textwidth]{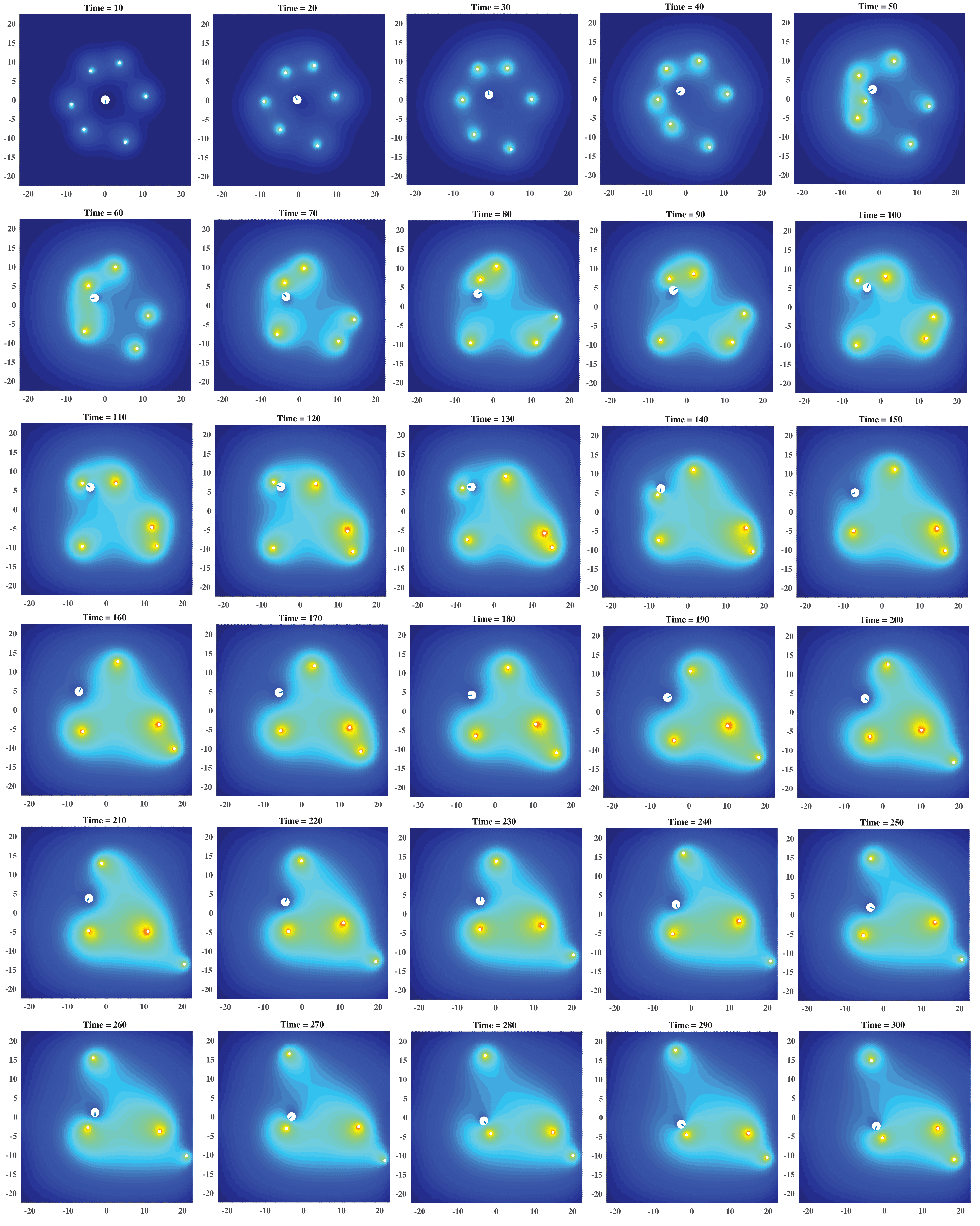}
  \caption{Time lapse snapshots of a typical simulation with 6 bacteria, $D = 1, \ \delta_J = 0.1$, no bacterial rheotaxis. The color scale
  is the same as main figure 3C.}  
  \label{Fig.mq7Ku6JH}
\end{figure}

\clearpage
\newpage
 \begin{figure}[h]
\center
  \includegraphics[width=0.95\textwidth]{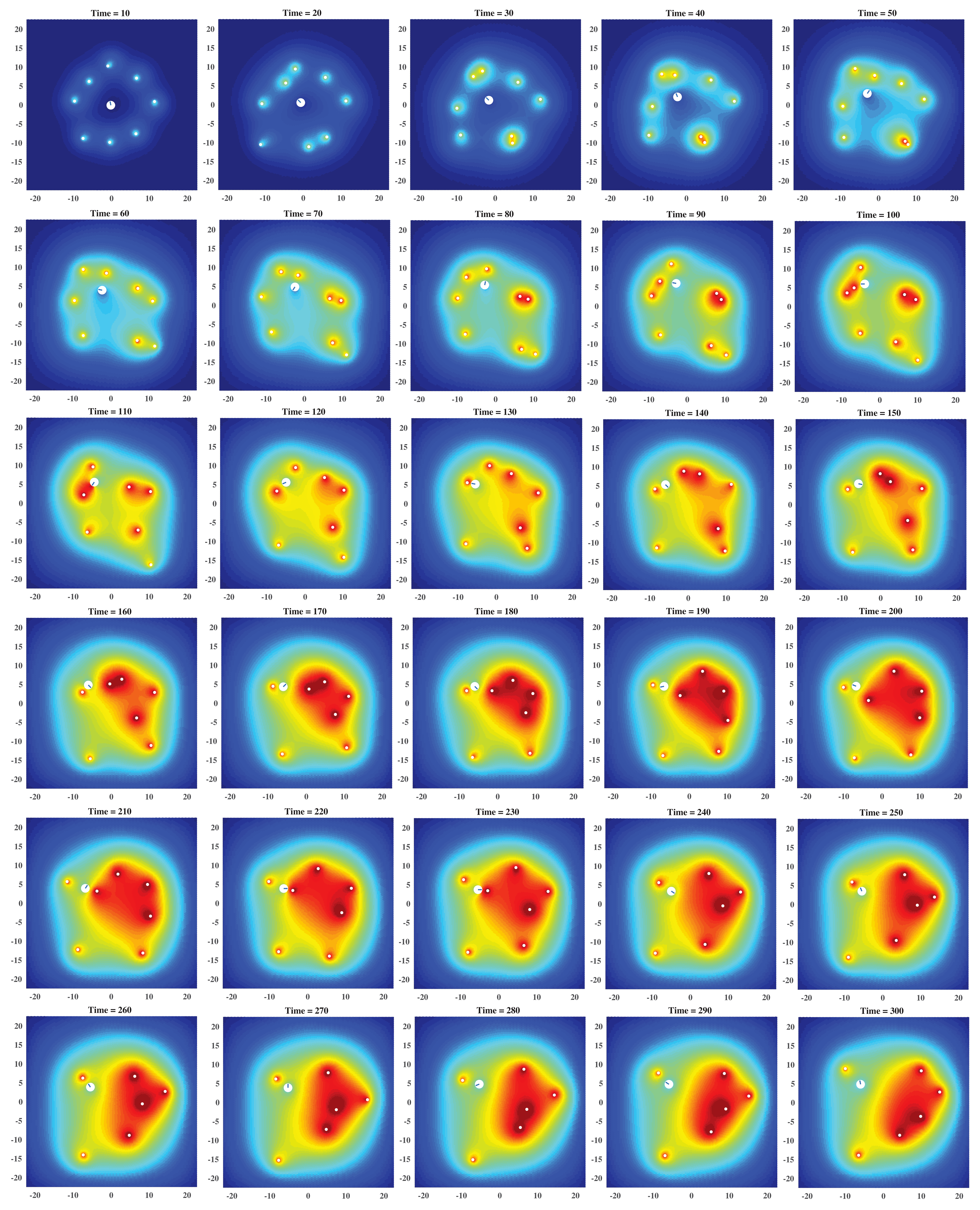}
  \caption{Time lapse snapshots of a typical simulation with 8 bacteria, $D = 1, \ \delta_J = 0.1$, no bacterial rheotaxis. The color scale
  is the same as main figure 3C.} 
  \label{Fig.bYRgUJ4m}
\end{figure}

\clearpage
\newpage
 \begin{figure}[h]
\center
  \includegraphics[width=0.95\textwidth]{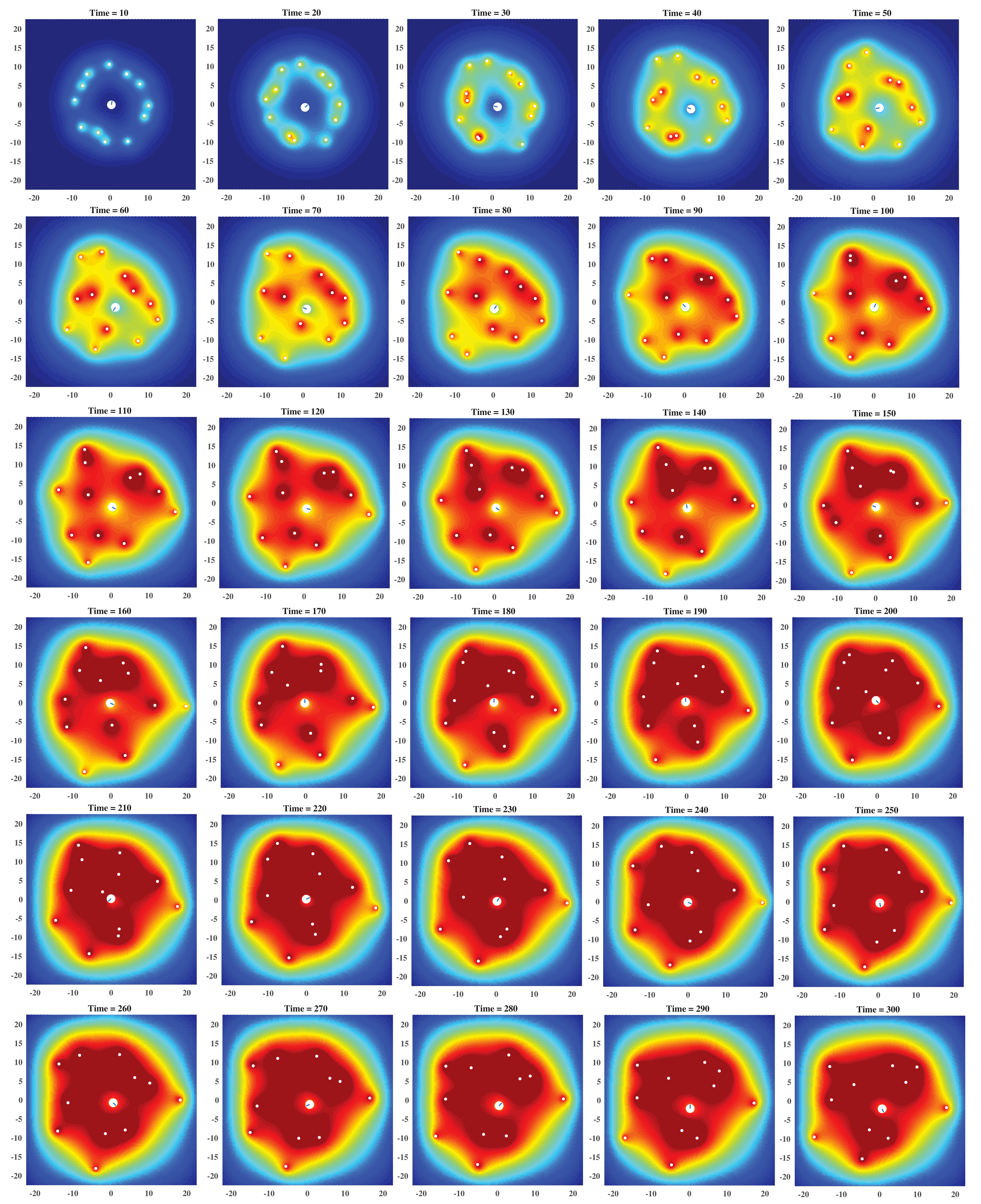}
  \caption{Time lapse snapshots of a typical simulation with 12 bacteria, $D = 1, \ \delta_J = 0.1$, no bacterial rheotaxis. The color scale
  is the same as main figure 3C.} 
    \label{Fig.ssI3XJjA}
\end{figure}



\section*{References}
\begin{thebibliography}{99} 

\bibitem{Delanoe2008Changes} Delanoë‐Ayari H, Iwaya S, Maeda YT, Inose J, Riviere C, Sano M, Rieu JP
2008
Changes in the magnitude and distribution of forces at different Dictyostelium developmental stages
{\it Cell. Motil. Cytoskeleton}
{\bf 65(4)} 314-31

\bibitem{Sokolov2010Swimming} Sokolov A, Apodaca MM, Grzybowski BA, Aranson IS
2010
Swimming bacteria power microscopic gears
{\it PNAS}
{\bf 107(3)} 969-74

\bibitem{Riedel2005Self} Riedel IH, Kruse K, Howard J
2005
A self-organized vortex array of hydrodynamically entrained sperm cells
{\it Science}
{\bf 309(5732)} 300-3

\bibitem{Lauga2006Swimming}
Lauga E, DiLuzio WR, Whitesides GM, Stone HA
2006
Swimming in circles: motion of bacteria near solid boundaries 
{\it Biophys. J.}
{\bf 90(2)} 400-12

\bibitem{Rafai2010Effective} Rafaï S, Jibuti L, Peyla P
2010
Effective viscosity of microswimmer suspensions
{\it Phys. Rev. Lett.}
{\bf 104(9)} 098102

\bibitem{Manahan2004Chemoattractant} Manahan CL, Iglesias PA, Long Y, Devreotes PN
2004
Chemoattractant signaling in Dictyostelium discoideum
{\it Annu. Rev. Cell Dev. Biol.}
{\bf 20} 223-53

\bibitem{howe2013amoebae} Howe JD, Barry NP, Bretscher MS
2013
How do amoebae swim and crawl?
{\it PLoS One}  
{\bf 8(9)} e74382

\bibitem{Bray2000Cell} Bray D
2000
{\it Cell movements: from molecules to motility}
(Garland Science)

\bibitem{Stewart1987molecular} Stewart RC, Dahlquist FW
1987
Molecular components of bacterial chemotaxis
{\it Chem. Rev.} 
{\bf 87(5)} 997-1025

\bibitem{Block1982Impulse} Block SM, Segall JE, Berg HC
1982
Impulse responses in bacterial chemotaxis
{\it Cell}   
{\bf 31(1)}
215-26

\bibitem{Berg1990bacterial} Berg HC
1990 
Bacterial microprocessing
{\it Cold Spring Harbor symposia on quantitative biology} 
(Cold Spring Harbor Laboratory Press)
p~539-545 

\bibitem{Paluch2005Cortical} Paluch E, Piel M, Prost J, Bornens M, Sykes C
2005
Cortical actomyosin breakage triggers shape oscillations in cells and cell fragments
{\it Biophys. J.}
{\bf 89(1)} 724-33

\bibitem{Tinevez2009Role} Tinevez JY, Schulze U, Salbreux G, Roensch J, Joanny JF, Paluch E
2009
Role of cortical tension in bleb growth
{\it PNAS}
{\bf 106(44)} 18581-6

\bibitem{Kay2002Chemotaxis} Kay RR
2002
Chemotaxis and cell differentiation in Dictyostelium
{\it Curr. Opin. Microbiol.}
{\bf 5(6)} 575-9

\bibitem{Kessin2001Dictyostelium} Kessin RH
2001
{\it Dictyostelium: evolution, cell biology, and the development of multicellularity}
(Cambridge University Press) 

\bibitem{Bonner2009Social} Bonner JT
2009
{\it The social amoebae: the biology of cellular slime molds}
(Princeton University Press)

\bibitem{Devreotes1988Chemotaxis} Devreotes PN, Zigmond SH
1988
Chemotaxis in eukaryotic cells: a focus on leukocytes and Dictyostelium
{\it Annu. Rev. Cell Biol}
{\bf 4(1)} 649-86

\bibitem{Bonner2015Cellular} Bonner JT
2015
{\it Cellular slime molds}
(Princeton University Press) 

\bibitem{Meinhardt1999Orientation} Meinhardt H
1999
Orientation of chemotactic cells and growth cones: models and mechanisms
{\it J. Cell Sci.}
{\bf 112(17)} 2867-74

\bibitem{Neilson2011Modeling} Neilson MP, Mackenzie JA, Webb SD, Insall RH
2011
Modeling cell movement and chemotaxis using pseudopod-based feedback
{\it SIAM J. Sci. Comput.}
{\bf 33(3)} 1035-57

\bibitem{Neilson2011Chemotaxis} Neilson MP, Veltman DM, van Haastert PJ, Webb SD, Mackenzie JA, Insall RH
2011
Chemotaxis: a feedback-based computational model robustly predicts multiple aspects of real cell behaviour
{\it PLoS Biol.}
{\bf 9(5)} e1000618

\bibitem{Hecht2011Activated} Hecht I, Skoge ML, Charest PG, Ben-Jacob E, Firtel RA, Loomis WF, Levine H, Rappel WJ
2011
Activated membrane patches guide chemotactic cell motility
{\it PLoS Comput. Biol.}
{\bf 7(6)} e1002044




\bibitem{Tang1995Excitation} Tang Y, Othmer HG
1995
Excitation, oscillations and wave propagation in a G-protein-based model of signal transduction in Dictyostelium discoideum
{\it Philos. Trans. R. Soc. Lond., B, Biol. Sci.}
{\bf 349(1328)} 179-95

\bibitem{Dallon1997Discrete} Dallon JC, Othmer HG
1997
A discrete cell model with adaptive signalling for aggregation of Dictyostelium discoideum
{\it Philos. Trans. R. Soc. Lond., B, Biol. Sci.}
{\bf 352(1351)} 391-417

\bibitem{Palsson2000Model} Palsson E, Othmer HG
2000
A model for individual and collective cell movement in Dictyostelium discoideum.  
{\it PNAS}
{\bf 97(19)} 10448-53

\bibitem{Dallon2004How} Dallon JC, Othmer HG
2004
How cellular movement determines the collective force generated by the Dictyostelium discoideum slug
{\it J. Theor. Biol.}
{\bf 231(2)} 203-22

\bibitem{Khamviwath2013Continuum} Khamviwath V, Hu J, Othmer HG
2013
A continuum model of actin waves in Dictyostelium discoideum
{\it PLoS One}
{\bf 8(5)} e64272

\bibitem{Cheng2016Model} Cheng Y, Othmer H
2016
A model for direction sensing in Dictyostelium discoideum: Ras activity and symmetry breaking driven by a $G_{\beta \gamma}$-mediated, $G_{\alpha 2}$-Ric8--dependent signal transduction network
{\it PLoS Comput. Biol.}
{\bf 12(5)} e1004900

\bibitem{Bretschneider2016Progress} Bretschneider T, Othmer HG, Weijer CJ
2016
Progress and perspectives in signal transduction, actin dynamics, and movement at the cell and tissue level: lessons from Dictyostelium
{\it Interface Focus}
{\bf 6(5)} 20160047

\bibitem{Dallon1998Continuum} Dallon JC, Othmer HG
1998
A continuum analysis of the chemotactic signal seen bydictyostelium discoideum
{\it J. Theor. Biol.}
{\bf 194(4)} 461-83

\bibitem{Wilhelm2007Magnetic} Wilhelm C, Riviere C, Biais N
2007
Magnetic control of Dictyostelium aggregation
{\it Phys. Rev. E}
{\bf 75(4)} 041906

\bibitem{Janmey2007Cell} Janmey PA, McCulloch CA
2007
Cell mechanics: integrating cell responses to mechanical stimuli
{\it Annu. Rev. Biomed. Eng.}
{\bf 9} 1-34

\bibitem{Riviere2007Signaling} Rivi\`ere C, Marion S, Guill\'en N, Bacri JC, Gazeau F, Wilhelm C
2007
Signaling through the phosphatidylinositol 3-kinase regulates mechanotaxis induced by local low magnetic forces in Entamoeba histolytica
{\it J. Biomech.}
{\bf 40(1)} 64-77

\bibitem{Decave2003Shear} D\'ecav\'e E, Rieu D, Dalous J, Fache S, Bréchet Y, Fourcade B, Satre M, Bruckert F
2003
Shear flow-induced motility of Dictyostelium discoideum cells on solid substrate
{\it J. Cell Sci.}
{\bf 116(21)} 4331-43

\bibitem{Holmes2012Comparison} Holmes WR, Edelstein-Keshet L
2012
A comparison of computational models for eukaryotic cell shape and motility
{\it PLoS Comput Biol.}
{\bf 8(12)} e1002793

\bibitem{van2011amoeboid} Van Haastert PJ
2011
Amoeboid cells use protrusions for walking, gliding and swimming
{\it PloS One} 
{\bf 6(11)} e27532

\bibitem{barry2010dictyostelium} Barry NP, Bretscher MS
2010
Dictyostelium amoebae and neutrophils can swim
{\it PNAS}
{\bf 107(25)} 1376-80

\bibitem{Franz2018Fat} Franz A, Wood W, Martin P
2018
Fat body cells are motile and actively migrate to wounds to drive repair and prevent infection
{\it Dev. Cell}
{\bf 44(4)} 460-70

\bibitem{wang2016computational} Wang Q, Othmer HG
2016
Computational analysis of amoeboid swimming at low Reynolds number
{\it J. Math. Biol.}
{\bf 72(7)} 1893-926.

\bibitem{Wu2015Amoeboid} Wu H, Thi\'ebaud M, Hu WF, Farutin A, Rafa{\"i} S, Lai MC, Peyla P, Misbah C
2015
Amoeboid motion in confined geometry
{\it Phys. Rev. E}
{\bf 92(5)} 050701

\bibitem{Wu2016Amoebid} Wu H, Farutin A, Hu WF, Thi\'ebaud M, Rafa{\"i} S, Peyla P, Lai MC, Misbah C
2016
Amoeboid swimming in a channel
{\it  Soft Matter}
{\bf 12(36)} 7470-84

\bibitem{Aoun2020Amoeboid} Aoun L, Farutin A, Garcia-Seyda N, Nègre P, Rizvi MS, Tlili S, Song S, Luo X, Biarnes-Pelicot M, Galland R, Sibarita JB
2020
Amoeboid swimming is propelled by molecular paddling in Lymphocytes
{\it Biophys. J.}
{\bf 119(6)} 1157-77

\bibitem{Dalal2020Amoeboid} Dalal S, Farutin A, Misbah C
2020
Amoeboid swimming in a compliant channel
{\it Soft Matter}
{\bf 16(6)} 1599-613


\bibitem{Farutin2013Amoeboid} Farutin A, Rafaï S, Dysthe DK, Duperray A, Peyla P, Misbah C
2013
Amoeboid swimming: a generic self-propulsion of cells in fluids by means of membrane deformations
{\it Phys. Rev. Lett.}
{\bf 111(22)} 228102

\bibitem{Bouffanais2010hydrodynamics} Bouffanais R, Yue DK
2010
Hydrodynamics of cell-cell mechanical signaling in the initial stages of aggregation
{\it Phys. Rev. E}
{\bf 81(4)} 041920

\bibitem{Campbell2017Computational} Campbell EJ, Bagchi P
2017
A computational model of amoeboid cell swimming. 
{\it Phys. Fluids}
{\bf 29(10)} 101902


\bibitem{Campbell2020Computational} Campbell EJ, Bagchi P
2020
A computational study of amoeboid motility in 3D: the role of extracellular matrix geometry, cell deformability, and cell–matrix adhesion
{\it Biomech Model Mechanobiol} 1-25


\bibitem{shapere1987self} Shapere A, Wilczek F
1987
Self-propulsion at low Reynolds number
{\it Phys. Rev. Lett}
{\bf 58(20)} 2051

\bibitem{shapere1989geometry} Shapere A, Wilczek F
1989
Geometry of self-propulsion at low Reynolds number
{\it J. Fluid Mech.}
{\bf 198} 557-85


\bibitem{avron2004optimal} Avron JE, Gat O, Kenneth O
2004
Optimal swimming at low Reynolds numbers  
{\it Phys. Rev. Lett}
{\bf 93(18)} 186001

\bibitem{Grossman1982Changes} Grossman N, Ron EZ, Woldringh CL
1982
Changes in cell dimensions during amino acid starvation of Escherichia coli
{\it J. Bacteriol.}
{\bf 152(1)} 35-41

\bibitem{Fu2012Bacterial} Fu HC, Powers TR, Stocker R
2012
Bacterial rheotaxis
{\it PNAS}  
{\bf 109(13)} 4780-5

\bibitem{pan2018g} Pan M, Neilson MP, Grunfeld AM, Cruz P, Wen X, Insall RH, Jin T
2018 
A G-protein-coupled chemoattractant receptor recognizes lipopolysaccharide for bacterial phagocytosis
{\it PLoS Biol.}
{\bf 16(5)} e2005754

\bibitem{cosson2008eat} Cosson P, Soldati T
2008
Eat, kill or die: when amoeba meets bacteria
{\it Curr. Opin. Microbiol.}  
{\bf 11(3)} 271-6

\bibitem{bottino2001computer} Bottino DC
2001
Computer simulations of mechanochemical coupling in a deforming domain: applications to cell motion
{\it Mathematical Models for Biological Pattern Formation}
(Springer, New York, NY) 
p~295-314

\bibitem{dillon2008single} Dillon R, Owen M, Painter K
2008
A single-cell-based model of multicellular growth using the immersed boundary method
{\it AMS Contemp. Math.}

\bibitem{borgers1987lagrangian} B{\"o}rgers C, Peskin CS
1987
A Lagrangian fractional step method for the incompressible Navier-Stokes equations on a periodic domain 
{\it J. Comput. Phys.}
{\bf 70(2)} 397-438

{\bf 466} 1-5

\bibitem{bae2010swimming} Bae AJ, Bodenschatz E
2010
On the swimming of Dictyostelium amoebae
{\it  PNAS}
{\bf 107(44)} E165-6




\bibitem{muskhelishvili2013some} Muskhelishvili NI
2013
{\it Some basic problems of the mathematical theory of elasticity}
(Springer Science \& Business Media)  

\bibitem{Cross1993Pattern} Cross MC, Hohenberg PC
1993
Pattern formation outside of equilibrium
{\it Rev. Mod. Phys.}
{\bf 65(3)} 851



\bibitem{Elliott2012Modelling} Elliott CM, Stinner B, Venkataraman C
2012
Modelling cell motility and chemotaxis with evolving surface finite elements
{\it J. R. Soc. Interface}
{\bf 9(76)} 3027-44


\bibitem{Jana2012Paramecium} Jana S, Um SH, Jung S
2012
Paramecium swimming in capillary tube
{\it Phys. Fluids}
{\bf 24(4)} 041901


\bibitem{Ledesma2013Enhanced} Ledesma-Aguilar R, Yeomans JM
2013
Enhanced motility of a microswimmer in rigid and elastic confinement
{\it Phys. Rev. Lett.}
{\bf 111(13)} 138101






\bibitem{Sokolov2007Concentration} Sokolov A, Aranson I S, Kessler J O and Goldstein R E
2007
Concentration dependence of the collective dynamics of swimming bacteria
{\it Phys. Rev. Lett.}
{\bf 98} 158102

\bibitem{Lushi2012Collective} Lushi E, Goldstein R E and Shelley M J 
2012 
Collective chemotactic dynamics in the presence of self-generated fluid flows 
{\it Phys. Rev. E}
{\bf 86} 040902

\bibitem{Lushi2018Nonlinear} Lushi E, Goldstein R E and Shelley M J
2018 
Nonlinear concentration patterns and bands in autochemotactic suspensions
{\it Phys. Rev. E}
{\bf 98} 052411

\bibitem{Nejad2019Chemotaxis} Nejad M R and Najafi A
2019
Chemotaxis mediated interactions can stabilize the hydrodynamic instabilities in active suspensions
{\it Soft Matter}
{\bf 11} 3248–55

\bibitem{Partridge2019Escherichia} Partridge J D, Nhu N T, Dufour Y S and Harshey R M
2019
Escherichia coli remodels the chemotaxis pathway for swarming
{\it mBio} 
{\bf 10} e00316–19

\bibitem{Ryan2019Role} Ryan SD
2019
Role of hydrodynamic interactions in chemotaxis of bacterial populations
{\it Phys. Biol.}
{\bf 17(1)} 016003

\bibitem{Xue2011Travelling} Xue C, Hwang HJ, Painter KJ, Erban R
2011
Travelling waves in hyperbolic chemotaxis equations.
{\it Bull. Math. Biol.}
{\bf 73(8)} 1695





\end{thebibliography}
\end{document}